\documentclass[12pt,aps,prd,onecolumn,notitlepage,superscriptaddress,preprintnumbers,     nofootinbib,floatfix,longbibliography]{revtex4-2}

\usepackage[T1]{fontenc}
\usepackage[utf8]{inputenc}
\usepackage[english]{babel}
\usepackage{amsmath,amsthm,amssymb,amsfonts,mathrsfs,amsbsy,bm}
\usepackage{tensor}
\usepackage{slashed}
\usepackage{esint}
\usepackage{capt-of}
\usepackage[a4paper, margin=1.4cm]{geometry}
\usepackage{cancel}
\usepackage{graphicx}
\usepackage{multirow}
\usepackage{array}
\usepackage{booktabs}
\usepackage{makecell}
\usepackage{xcolor}
\colorlet{BLUE}{blue}
\usepackage{tikz}
\usetikzlibrary{quotes,angles,arrows,decorations.markings,decorations.pathmorphing}
\usepackage{hyperref}
\hypersetup{colorlinks=true,breaklinks=true,citecolor=blue,linkcolor=[rgb]{0,0.5,0.9},urlcolor=blue}
\graphicspath{{figures/}}

\newcommand{\be}{\begin{equation}}
\newcommand{\ee}{\end{equation}}
\newcommand{\Be}{\begin{eqnarray}}
\newcommand{\Ee}{\end{eqnarray}}

\newcommand{\mincir}{\raise-3.truept\hbox{\rlap{\hbox{$\sim$}}\raise4.truept\hbox{$<$}\ }}
\newcommand{\magcir}{\raise-3.truept\hbox{\rlap{\hbox{$\sim$}}\raise4.truept\hbox{$>$}\ }}

\providecommand{\U}[1]{}
\newcommand{\ie}{\begin{equation}}
\newcommand{\fe}{\end{equation}}
\newcommand{\se}{\begin{eqnarray}}
\newcommand{\ff}{\end{eqnarray}}

\begin{document}
\emergencystretch=4em


\title{Light bending and observational bounds in dyonic Kalb–Ramond gravity}

\author{A. A. Ara\'{u}jo Filho}
\email{dilto@fisica.ufc.br}
\affiliation{Departamento de F\'isica, Universidade Federal da Para\'iba, Caixa Postal 5008, 58051--970, Jo\~ao Pessoa, Para\'iba, Brazil.}
\affiliation{Departamento de F\'isica, Universidade Federal de Campina Grande, Caixa Postal 10071, 58429--900 Campina Grande, Para\'iba, Brazil.}
\affiliation{Center for Theoretical Physics, Khazar University, 41 Mehseti Street, Baku, AZ-1096, Azerbaijan.}


\date{\today}


\begin{abstract}

We investigate the weak and strong gravitational lensing of a dyonic black hole in Kalb--Ramond gravity. After fixing the asymptotic normalization, we distinguish the effective charge governing the local null trajectories from the global angular identification associated with Lorentz--symmetry breaking. In the weak--deflection regime, we evaluate the Gaussian curvature explicitly and apply the Gauss--Bonnet theorem with a perturbed ray boundary, recovering the complete second order mass contribution. Independent calculations based on the orbit equation, the turning point integral, and Fermat's principle reproduce the same bending angle. In the strong--deflection regime, Tsukamoto's method yields closed expressions for both strong--deflection coefficients, with exact charge dependence. We derive finite--distance lens equations and the associated image positions, magnifications, flux ratios, and differential arrival times, retaining the physical winding condition. Shadow sizes inferred from observations of Sgr~A$^*$ and M87$^*$ yield conditional charge bounds, complemented by an estimate from S2 precession. We further construct combinations of strong--lensing observables that separate the effective charge from the conical parameter within the model: angular separation and relative brightness remove the explicit winding dependence, while a timing ratio incorporating an independent distance estimate isolates the conical deformation.

\end{abstract}

\maketitle

\newpage

\tableofcontents


\section{Introduction}
\label{sec:introduction}

Gravitational lensing relates the geometry of a compact object to the angular positions, relative brightness, and arrival times of its images~\cite{Perlick2004,PerlickReview2004}. Rays passing far from the lens probe the asymptotic gravitational field, whereas trajectories approaching an unstable circular null orbit undergo large deflections and may complete several windings before reaching the observer. These regimes give rise to complementary information about the same spacetime. The horizon scale images of M87$^*$ and Sgr~A$^*$ obtained by the Event Horizon Telescope have supplied observational scales against which predictions for photon capture can be assessed~\cite{EHTM872019,EHTSgrI2022,EHTSgrVI2022}. Measurements of stellar motion near the Galactic center provide independent information on the central mass, distance, and relativistic orbital precession, which allows null and timelike probes to be considered together~\cite{GRAVITY2020,GRAVITY2022}.

One motivation for extending these tests beyond general relativity is the possibility that a tensor vacuum spontaneously breaks local Lorentz symmetry~\cite{Kostelecky1989,Kostelecky2004}. The Kalb--Ramond field, originally introduced in the description of interacting strings~\cite{Kalb1974}, supplies an antisymmetric two form realization of this mechanism. Its nonvanishing vacuum expectation value can select a preferred local orientation while the underlying action remains covariant. The dynamics and gravitational consequences of such antisymmetric backgrounds have been examined in Refs.~\cite{Higashijima2001,Altschul2010,Maluf2018}.

Black hole solutions with a background Kalb--Ramond field have developed from neutral configurations to electrically charged settings and more general exact families~\cite{Lessa2020,Yang2023,Duan2024,Liu2024,Liu2025,AraujoFilho:2025jcu}. Recent constructions have incorporated two independent curvature couplings~\cite{Yang2026} and simultaneous electric and magnetic charges~\cite{Lin2026}. Their phenomenology has been investigated through shadows, quasinormal modes, geodesics, scattering, evaporation~\cite{Ara2024,AraCharged2025}, within the context of neutrino physics \cite{Shi:2025xkd,Shi:2025rfq} and quasi periodic oscilations \cite{Rodrigues:2026ofx}. Additional matter sectors broaden this setting: it was obtained and examined global monopole geometries~\cite{BelchiorMonopole2025}, while ModMax electrodynamics has been considered in studies of particle dynamics, perturbations, and black hole thermodynamics~\cite{AhmedModMax2025,SekhmaniPhantom2025,SucuBranches2026}. More recently, it was incorporated a surrounding perfect fluid dark matter distribution~\cite{BelchiorModMax2026}.

Recent work beyond spherical black holes also illustrates why the interpretation of deformation parameters requires care. In a localized Kalb--Ramond wormhole, there was a construction of a slow rotation configuration with well defined asymptotic charges but a fractional quadrupolar tail that obstructs the usual smooth multipole construction~\cite{MurodovWormhole2026}. For a Newman--Janis generated rotating charged geometry, the analysis of massive neutral Dirac quasibound states showed that spectral trends can change when the asymptotic normalization and the quantities held fixed in a parameter scan are altered~\cite{MurodovDirac2026}. In addition, a study proposed a scaling in a power law rotating Kalb--Ramond background further found that the inferred deformation depended strongly on the magnetic flux prescription, with the adopted jet power proxies failing to determine it independently~\cite{MurodovBZ2026}. The investigation within two minimal couplings has aslso been reported \cite{Rahmatov2026}.

In the weak--deflection limit, the Gauss--Bonnet approach of Gibbons and Werner expresses light bending through the optical curvature and the boundary geometry of a suitable integration domain~\cite{GibbonsWerner2008}; its extension to stationary spacetimes was developed by Werner~\cite{Werner2012}. Finite distance formulations subsequently made the source and observer positions explicit~\cite{Ishihara2016,Ishihara2017,OnoAsada2019}, while expansions of the lens equation established how corrections to the bending angle enter image positions, magnifications, and time delays~\cite{KeetonPetters2005,KeetonPetters2006}. Beyond the leading mass term, the integration domain must fundamentaly be expanded with the curvature. A first order correction to the ray boundary contributes to the second order deflection, as implemented in recent applications~\cite{AraNoncomm2025,Jha2025}. This dependence on the perturbed trajectory makes comparisons with the orbit equation, the turning point integral, and Fermat's principle useful checks of the complete expansion~\cite{Sereno2004,IyerPetters2007}.

Weak lensing in Kalb--Ramond backgrounds has been investigated for neutral and charged geometries, including propagation through plasma and asymptotically (anti--)de Sitter configurations~\cite{Atamurotov2022,Pantig2025,PantigOvgun2025,Filho:2024tgy} and noncommunicative scenarios \cite{AraujoFilho:2025huk}. It was studied charged black hole photon dynamics using geometrical and perturbative methods~\cite{SucuSakalli2025}, whereas there existed a examination of the deflection and magnification in the presence of cosmic strings and string clouds~\cite{AhmedStrings2025}. Further developments include axion--plasmon effects in Ricci coupled backgrounds~\cite{Ovgun2025} and chromatic weak lensing with two Lorentz--violating curvature couplings~\cite{Ovgun2026}.

In the strong--deflection limit, the bending angle develops a logarithmic divergence as the impact parameter approaches its critical value. The relativistic image construction of Virbhadra and Ellis and the analytical framework developed by Bozza relate this behavior to observable image sequences~\cite{VirbhadraEllis2000,Bozza2001,Bozza2002}. Tsukamoto's reformulation shows a self consistent extraction of the divergent and regular contributions for static, spherically symmetric geometries~\cite{Tsukamoto2017}, with charged lenses supplying useful analytical results~\cite{Eiroa2002,TsukamotoGong2017}. Finite source distances and differential arrival times furnish additional information beyond the limiting image position~\cite{BozzaScarpetta2007,BozzaMancini2004}. Within Kalb--Ramond gravity, there are also a study relativistic images in a global monopole background~\cite{BelchiorLensing2025},  charged configurations with nonlinear electrodynamics~\cite{PereiraKR2026}, and deflection regimes in the presence of an anisotropic fluid~\cite{SekhmaniAnisotropic2026}.

The observational interpretation of these results requires a distinction between a geometrical critical curve, the associated shadow, and the bright emission structure reconstructed from interferometric data~\cite{Gralla2019,PerlickTsupko2022}. Constraints inferred from shadow sizes depend on the adopted mass and distance information and on the relation between the emission region and the capture boundary~\cite{EHTSgrVI2022,TsukamotoKase2024}. Recent work on rotating Kalb--Ramond geometries has explored this connection with M87$^*$ and Sgr~A$^*$~\cite{SekhmaniRotating2025}. Complementary proposals based on the interferometric signature and temporal correlations of photon rings aim to access information beyond an overall angular diameter~\cite{Johnson2020,Hadar2021,Ayzenberg2025,LupsascaBHEX2024}.

In this work, we investigate the weak and strong gravitational lensing of minimally coupled test radiation by the dyonic Kalb--Ramond black hole obtained in Ref.~\cite{Lin2026} (with very recent gravitational implications \cite{AraujoFilho:2026tsc}). The coexistence of electric and magnetic charges is relevant because their contributions to the metric carry different dependences on the Lorentz--violating coupling. After fixing the asymptotic normalization, the local equatorial null trajectories take a Reissner--Nordstr\"om form with an effective charge, while the global angular identification retains a conical deformation. We use this separation to organize the calculation from the weak--field expansion to the physical winding condition of the relativistic images. The analysis combines an explicit optical curvature calculation and independent weak deflection derivations with closed strong--deflection coefficients, finite distance lens equations, and angular, photometric, and timing observables. We then translate shadow size intervals and stellar precession information into conditional bounds within the model. Finally, we construct combinations of strong lensing observables that separate the effective charge from the conical parameter at the stated approximation order, while retaining the degeneracy between the underlying electric and magnetic charges.

The paper is organized as follows. Section~\ref{sec:model} summarizes the dyonic solution and its admissible configuration. Sections~\ref{sec:weak} and~\ref{sec:bounds} develop the weak deflection analysis and its observational implications. Sections~\ref{sec:strong} and~\ref{sec:strongbounds} present the strong--deflection calculation and the corresponding bounds and reconstruction relations. Our conclusions are given in Sec.~\ref{sec:conclusion}.


\section{The dyonic black hole and the general features}
\label{sec:model}

An antisymmetric tensor field provides a natural setting in which a gravitational background can acquire a preferred local orientation. The Kalb--Ramond field, originally introduced in the description of interacting strings~\cite{Kalb1974}, realizes this possibility through a nonvanishing vacuum expectation value. Spontaneous Lorentz symmetry breaking in tensor theories has been examined both within string inspired models and in the gravitational sector of the Standard Model Extension~\cite{Kostelecky1989,Kostelecky2004,Higashijima2001,Altschul2010}. Its approach for an antisymmetric two form also modifies the propagating degrees of freedom, as discussed in Ref.~\cite{Maluf2018}. Here, we consider the dyonic solution obtained by Lin, Liu, and Liu~\cite{Lin2026}, retaining the ingredients needed for the subsequent lensing analysis.

In geometrized units, $G = c = 1$, and with metric signature $(-,+,+,+)$, the action is written as
\begin{align}
S = {}&\int \mathrm{d}^4x\sqrt{-g} \left[\frac{1}{2\kappa}
 \left(R - 2\Lambda + \xi B^{\mu\rho}B^{\nu}{}_{\rho}R_{\mu\nu}\right)
 - \frac{1}{12}H_{\mu\nu\rho}H^{\mu\nu\rho}-V(X)
 + \mathcal L_{\mathrm{em}}\right],\label{eq:action}\\
\mathcal L_{\mathrm{em}} = {}&-\frac{1}{2\kappa}\left[
 F_{\mu\nu}F^{\mu\nu}
 + \gamma_1\left(B^{\mu\nu}F_{\mu\nu}\right)^2
 + \gamma_2 B_{\mu\nu}B^{\mu\nu}F_{\rho\sigma}F^{\rho\sigma}\right],
 \label{eq:emlag}
\end{align}
where $\kappa = 8\pi$, $F_{\mu\nu} = \partial_\mu A_\nu - \partial_\nu A_\mu$, and
\begin{equation}
H_{\mu\nu\rho} = \partial_\mu  B_{\nu\rho} + \partial_\nu B_{\rho\mu}
  + \partial_\rho B_{\mu\nu}, \qquad
X = B_{\mu\nu}B^{\mu\nu} + b_0^2.
\label{eq:KRdefinitions}
\end{equation}
The constant $b_0$ specifies the vacuum norm; naturally, it is distinct from the impact parameter introduced below. The parameter $\xi$ controls the nonminimal curvature coupling, whereas $\gamma_1$ and $\gamma_2$ determine two different interactions with the electromagnetic field. Such couplings extend the neutral and electrically charged black hole configurations previously constructed in Kalb--Ramond gravity~\cite{Lessa2020,Yang2023,Duan2024,Liu2024,Liu2025}. More general charged part with two independent curvature couplings have recently been obtained as well~\cite{Yang2026}; those additional curvature couplings are not included in Eq.~\eqref{eq:action}.

We restrict the analysis to $\Lambda=0$ and to the vacuum of the quadratic potential $V(X)=\lambda X^2/2$, for which $X=0$ and $V=V'=0$. A static, spherically symmetric configuration may be expressed as
\begin{equation}
\mathrm{d}s^2 = - F(r) \mathrm{d}t^2 + G(r) \mathrm{d}r^2 + r^2\mathrm{d}\Omega^2,\qquad
B_{tr} = \frac{b_0}{\sqrt{2}}\sqrt{F(r)G(r)},\qquad
A = \psi(r)\mathrm{d}t+p\cos\vartheta\,\mathrm{d}\phi,
\label{eq:ansatz}
\end{equation}
where $\mathrm{d}\Omega^2 = \mathrm{d}\vartheta^2+\sin^2\vartheta\,\mathrm{d}\phi^2$. Notice that  this vacuum satisfies $B_{\mu\nu}B^{\mu\nu} = - b_0^2$ and $H_{\mu\nu\rho} = 0$. The magnetic potential is understood in the usual overlapping gauge patches, and $p$ denotes its monopole charge parameter. In addition, variation with respect to $A_\mu$ gives the modified Maxwell equation
\begin{equation}
\nabla_\mu\mathcal D^{\mu\nu} = 0, \qquad
\mathcal D^{\mu\nu} = \left(1 + \gamma_2B_{\alpha\beta}B^{\alpha\beta}\right)F^{\mu\nu}
 +\gamma_1 \left(B^{\alpha\beta} F_{\alpha\beta}\right)B^{\mu\nu},
\label{eq:maxwell}
\end{equation}
which implies that the electric charge $Q$ is the conserved, as we should expect.

The difference between the radial and temporal gravitational equations fixes $FG$ to a constant, which can be set to unity by a constant rescaling of $t$. On the genuinely dyonic setting, compatibility of the remaining gravitational and two form equations imposes that~\cite{Lin2026}
\begin{equation}
\ell = \frac{\xi b_0^2}{2}, \qquad
\gamma_1 + \gamma_2=\frac{\xi}{2}, \qquad
\gamma_2 = \frac{\xi}{2b_0^2\xi-2}.
\label{eq:couplings}
\end{equation}
The radial equations reduce to follwoing forms
\begin{equation}
\frac{\mathrm{d}}{\mathrm{d}r}\!\left[r^2(1-\ell)\psi'(r)\right] = 0,
\qquad
\frac{\mathrm{d}}{\mathrm{d}r}\left[rf(r)\right] = a - \frac{\mathcal C}{r^2},
\qquad
a = \frac{1}{1 - \ell},\qquad
\mathcal C = \frac{Q^2}{(1 - \ell)^2} + \frac{p^2}{1 - 2\ell}.
\label{eq:reduced}
\end{equation}
Choosing the potential to vanish at infinity, and fixing the electric flux orientation so that $r^2(1-\ell)\psi' = -Q$, we obtain
\begin{equation}
\mathrm{d}s^2 = -f(r)\mathrm{d}t^2 + \frac{\mathrm{d}r^2}{f(r)} + r^2\mathrm{d}\Omega^2, 
\qquad 
f(r) = a - \frac{2M}{r} + \frac{\mathcal C}{r^2},
 \qquad 
\psi(r) = \frac{Q}{(1 - \ell)r}.
\label{eq:solution}
\end{equation}
The electric and magnetic contributions have the same radial falloff, but their dependence on $\ell$ is different. The dyonic Reissner--Nordstr\"om geometry follows when $\ell\to0$, while $p\to0$ gives the electrically charged sector of Ref.~\cite{Duan2024}.

For definiteness, we work on the part that continuously connect to Einstein--Maxwell theory with $
\ell< 1/2$, $M>0$, and $M^2\geq a\,\mathcal{C}$.
These conditions ensure $a>0$, nonnegative charge contributions, and an event horizon. They also make the background electromagnetic coefficients $1 - \gamma_2\,b_0^2 = (1 - \ell)/(1-2\ell)$ and $1 - (\gamma_1 + \gamma_2)\,b_0^2 = 1 - \ell$ positive; they are not a substitute for a stability analysis of the full coupled theory. The horizon radii are
\begin{equation}
r_\pm = \frac{M\pm\sqrt{M^2 - a\,\mathcal C}}{a}.
\label{eq:horizons}
\end{equation}
The parameter $M$ is the integration constant appearing in the metric. Since $f(\infty)=a$, it should not be assigned the usual asymptotically Minkowskian mass normalization without specifying the asymptotic clock and radial scale.


\section{Weak gravitational lensing}
\label{sec:weak}

The weak deflection regime probes the exterior geometry at distances large compared with the gravitational radius. Optical signatures of Kalb--Ramond backgrounds have been investigated through shadows, null trajectories, and weak and strong gravitational lensing~\cite{Ara2024,AraCharged2025,Pantig2025}. Related calculations for charged and nonlinear electrodynamic geometries provide useful comparisons~\cite{Eiroa2002,Sereno2004,Fu2021,AraJCAP2025}. In the present case, the dyonic charge combination must be treated together with the nonstandard asymptotic geometry. In other words, we fix the angular convention before calculating the bending to obtain the same weak--field coefficients from optical curvature, the orbit equation, a turning point integral, and Fermat's principle.

\subsection{Asymptotic normalization and the optical geometry}
\label{subsec:normalization}

Initially, let us introduce
\begin{equation}
T = \sqrt a\,t, \qquad R = \frac{r}{\sqrt a},\qquad
\mu = \frac{M}{a^{3/2}}, \qquad q^2 = \frac{\mathcal C}{a^2},
\label{eq:normalization}
\end{equation}
the line element becomes
\begin{equation}
\mathrm{d}s^2 = -A(R)\mathrm{d}T^2 + \frac{\mathrm{d}R^2}{A(R)} + aR^2\mathrm{d}\Omega^2,
\qquad A(R) = 1 - \frac{2\mu}{R} + \frac{q^2}{R^2}.
\label{eq:normalizedmetric}
\end{equation}
Here, $\mu$ is the coefficient determining the normalized Newtonian potential. Since it is well konwn, the spherical symmetry allows each ray to be placed in the equatorial plane. On this plane, the unwrapped angle $\varphi=\sqrt a\,\phi$ gives
\begin{equation}
\left.\mathrm{d}s^2\right|_{\vartheta = \pi/2}
= - A(R)\mathrm{d}T^2 + A^{-1}(R)\mathrm{d}R^2 + R^2\mathrm{d}\varphi^2.
\label{eq:equatorialRN}
\end{equation}
This is locally the Reissner--Nordstr\"om equatorial metric. The global identification remains $\varphi\sim\varphi+2\pi\sqrt a$; however, it is worthy to be mentioined that the full four dimensional metric is not transformed into the Reissner--Nordstr\"om spacetime. This distinction is central to lensing in an asymptotically conical geometry~\cite{Perlick2004,Ovgun2026}.

The parameters controlling the local orbit are
\begin{equation}
\mu = M(1 - \ell)^{3/2}, \qquad
q^2 = Q^2 + \frac{(1-\ell)^2}{1 - 2\ell}p^2 
= Q^2 + p^2 + \frac{\ell^2}{1 - 2\ell}p^2.
\label{eq:effectiveparameters}
\end{equation}
In particular, the electric contribution to $q^2$ is independent of $\ell$, whereas the additional magnetic contribution begins at order $\ell^2$. Notice that such a cancellation would be hidden by expanding the original metric before normalizing its asymptotic coordinates.

Let $E = f\dot t$ and $L = r^2\dot\phi$ denote the original conserved quantities. Their normalized counterparts are $\mathcal E=A\dot T=E/\sqrt a$ and $J=R^2\dot\varphi=L/\sqrt a$, so that
\begin{equation}
b = \frac{J}{\mathcal E} = \frac{L}{E}.
\label{eq:impact}
\end{equation}
The impact parameter $b$ is the perpendicular separation on the asymptotic covering plane. For a scattering ray with both endpoints at infinity, we define
\begin{equation}
\widehat\alpha = \Delta\varphi - \pi, \qquad
\widehat\alpha_{\phi}^{\mathrm{bg}}
 = \Delta\phi - \frac{\pi}{\sqrt a}
 = \frac{\widehat\alpha}{\sqrt a},\qquad
\widehat\alpha_{\phi}^{\mathrm{coord}}
 = \Delta\phi - \pi
 = \pi\left(\frac{1}{\sqrt a}- 1\right)
  + \frac{\widehat\alpha}{\sqrt a}.
\label{eq:angleconventions}
\end{equation}
In this manner, $\widehat\alpha$ measures bending relative to the straight ray on the unwrapped asymptotic cone. The constant term in $\widehat\alpha_{\phi}^{\mathrm{coord}}$ describes the global angular identification and survives when $M = Q = p = 0$. It will enter the source--observer construction separately.

Throughout the weak--field expansion, we count
\begin{equation}
\frac{\mu}{b} = O(\epsilon),\qquad
\frac{q^2}{b^2} = O(\epsilon^2),\qquad \epsilon\ll1,
\label{eq:counting}
\end{equation}
while keeping the dependence on $\ell$ exact unless stated otherwise. This is a weak--field expansion for a charged black hole with fixed $q/\mu$. The exterior scattering part requires $b>b_{\mathrm{ph}}$, where
\begin{equation}
R_{\mathrm{ph}} = \frac{3\mu+\sqrt{9\mu^2-8q^2}}{2}, \qquad 
b_{\mathrm{ph}} = \frac{R_{\mathrm{ph}}}{\sqrt{A(R_{\mathrm{ph}})}}.
\label{eq:critical}
\end{equation}
Only the domain $b\gg\mu$ is used below; Eq.~\eqref{eq:critical} fixes the scattering segment without invoking a strong--deflection expansion.


\subsection{The Gauss--Bonnet calculation}
\label{subsec:GB}

For an equatorial null trajectory, Eq.~\eqref{eq:equatorialRN} defines the optical metric
\begin{equation}
\mathrm{d}T^2 = \mathrm{d}\sigma^2 = \frac{\mathrm{d}R^2}{A^2(R)} + \frac{R^2}{A(R)}\mathrm{d}\varphi^2
 = \mathrm{d}R_*^2+\mathcal F^2(R_*)\mathrm{d}\varphi^2,
\qquad \mathrm{d}R_* = \frac{\mathrm{d}R}{A},\qquad \mathcal F = \frac{R}{\sqrt A}.
\label{eq:opticalmetric}
\end{equation}
Its Gaussian curvature follows directly from $K=-\mathcal F^{-1}\mathrm{d}^2\mathcal F/\mathrm{d}R_*^2$. In terms of the lapse function,
\begin{equation}
K = \frac{AA''}{2} - \frac{(A')^2}{4},\qquad
A' = \frac{2\mu}{R^2} - \frac{2q^2}{R^3},\qquad
A'' = -\frac{4\mu}{R^3} + \frac{6q^2}{R^4},
\label{eq:Kderivatives}
\end{equation}
so that
\begin{equation}
K(R) = - \frac{2\mu}{R^3}
  + \frac{3(\mu^2+q^2)}{R^4}
  - \frac{6\mu q^2}{R^5}
  + \frac{2q^4}{R^6}, \qquad
\mathrm{d}S = \frac{R}{A^{3/2}}\,\mathrm{d}R\, \mathrm{d}\varphi.
\label{eq:curvatureexact}
\end{equation}
Notice that both expressions are exact. The negative leading mass term produces attractive bending, while the leading charge term reduces it. The optical curvature is distinct from the four dimensional Ricci scalar and from the curvature of a spatial slice.

For comparison, the optical metric defined with the original time coordinate is $\mathrm{d}t^2 = f^{-2}\mathrm{d}r^2 + r^2 f^{-1}\mathrm{d}\phi^2$, with
\begin{equation}
K_t(r) = - \frac{2aM}{r^3}
 + \frac{3(M^2 + a\mathcal C)}{r^4}
  - \frac{6M\mathcal C}{r^5} + \frac{2\mathcal C^2}{r^6},
\qquad \mathrm{d}S_t = \frac{r}{f^{3/2}}\,\mathrm{d}r\,\mathrm{d}\phi.
\label{eq:originalcurvature}
\end{equation}
The constant clock rescaling gives $K=K_t/a$ and $\mathrm{d}S = a\,\mathrm{d}S_t$ on the same domain, leaving $K\,\mathrm{d}S$ invariant.

With all these preliminaries, we can properly apply the Gauss--Bonnet construction of Gibbons and Werner~\cite{GibbonsWerner2008}, whose optical--geometric extension also underlies rotating and dispersive calculations~\cite{Werner2012,CrisnejoGallo2018}. Let $\mathcal D$ be the simply connected region outside the ray, bounded by the ray and a large circular arc. The theorem gives
\begin{equation}
\iint_{\mathcal D}K\,\mathrm{d}S + \oint_{\partial\mathcal D}\kappa_g\,\mathrm{d}\sigma 
 + \sum_i\Theta_i=2\pi.
\label{eq:GBtheorem}
\end{equation}
The ray is an optical geodesic, so its geodesic curvature vanishes. For the circular boundary in the original coordinates,
\begin{equation}
\left.\kappa_g\,\mathrm{d}\sigma_t\right|_{r=\mathrm{const}}
 = \left(\sqrt f-\frac{rf'}{2\sqrt f}\right)\mathrm{d}\phi
\longrightarrow\sqrt a\,\mathrm{d}\phi = \mathrm{d}\varphi.
\label{eq:boundary}
\end{equation}
The asymptotic jump angles sum to $\pi$. Thereby,
\begin{equation}
\widehat\alpha = - \iint_{\mathcal D_\infty}K\,\mathrm{d}S, 
\qquad 
\widehat\alpha_\phi^{\mathrm{coord}}
=\pi\left(a^{-1/2} - 1\right)
  - \frac{1}{\sqrt a}\iint_{\mathcal D_\infty}K_t\,\mathrm{d}S_t.
\label{eq:GBangle}
\end{equation}
Replacing the limiting measure in Eq.~\eqref{eq:boundary} by $\mathrm{d}\phi$ would remove the conical boundary contribution~\cite{Ovgun2026}. To determine the bending through second order, we also maintain the first gravitational correction to the ray that bounds the integration domain. Following the trajectory expansion employed in Eq.~(5.5) of Ref.~\cite{AraNoncomm2025} and in Ref.~\cite{Jha2025}, we write, in the normalized coordinates,
\begin{equation}
U_\gamma(\varphi) \equiv \frac{1}{\widetilde R(\varphi)}
= \frac{\sin\varphi}{b}
 + \frac{\mu(1-\cos\varphi)^2}{b^2}
 + O\!\left(\frac{\mu^2}{b^3},\frac{q^2}{b^3}\right).
\label{eq:GBorbit}
\end{equation}
Here, let us remember that $\mu = M(1 - \ell)^{3/2}$ is the mass parameter appearing in the normalized lapse, as we have introduced before. The incoming asymptote is fixed by $U_\gamma(0)=0$ and $U_\gamma'(0) = 1/b$. In the original radial and angular coordinates, the same trajectory is
\begin{equation}
\frac{1}{\widetilde r(\phi)}
=\frac{\sin(\sqrt a\,\phi)}{\sqrt a\,b}
 +\frac{M[1 - \cos(\sqrt a\,\phi)]^2}{a^2b^2}
 +O\!\left(\frac{\epsilon^2}{\sqrt a\,b}\right),
\qquad \widetilde r = \sqrt a\,\widetilde R.
\label{eq:GBoriginalorbit}
\end{equation}
Although it is straightforward to mention,  for $\ell=0$, Eq.~\eqref{eq:GBoriginalorbit} reduces to $\widetilde r^{-1} = \sin\phi/b + M(1 - \cos\phi)^2/b^2$ at the retained order.

The curvature density must be expanded to the same accuracy. Eq.~\eqref{eq:curvatureexact} gives
\begin{equation}
K\,\mathrm{d}S = \left[ -\frac{2\mu}{R^2}
 + \frac{3(q^2 - \mu^2)}{R^3}
 + O\!\left(\frac{\mu^3}{R^4},\frac{\mu q^2}{R^4}\right)\right]
 \mathrm{d}R\,\mathrm{d}\varphi.
\label{eq:curvaturedensity}
\end{equation}
Since the leading curvature density is linear in $\mu$, the first order displacement in Eq.~\eqref{eq:GBorbit} generates a second order contribution to the deflection. Terms proportional to $\mu^2/b^3$ and $q^2/b^3$ in the orbit would enter the curvature integral only at third order. Likewise, replacing the outgoing endpoint $\pi+\widehat\alpha$ by $\pi$ affects the result only at third order. With $\widetilde R(\varphi)$ as the lower radial boundary, it reads
\begin{align}
\widehat\alpha
={}& - \int_0^\pi\int_{\widetilde R(\varphi)}^\infty
 K\frac{R}{A^{3/2}}\,\mathrm{d}R\,\mathrm{d}\varphi
 + O(\epsilon^3)
= \int_0^\pi\left[
 2\mu U_\gamma + \frac32(\mu^2 - q^2)U_\gamma^2
 \right]\mathrm{d}\varphi + O(\epsilon^3)\nonumber\\
= {}& \frac{2\mu}{b}\int_0^\pi\sin\varphi\,\mathrm{d}\varphi
 +\frac{2\mu^2}{b^2}\int_0^\pi(1 - \cos\varphi)^2\,\mathrm{d}\varphi
+ \frac{3(\mu^2 - q^2)}{2b^2}\int_0^\pi\sin^2\varphi\,\mathrm{d}\varphi
 +O(\epsilon^3)\nonumber\\
={}&\frac{4\mu}{b} + \frac{3\pi\mu^2}{b^2}
 +\frac{3\pi(\mu^2 - q^2)}{4b^2} +O(\epsilon^3) 
={}\frac{4\mu}{b}
 +\frac{\pi(15\mu^2 - 3q^2)}{4b^2} +O(\epsilon^3).
\label{eq:deflection}
\end{align}

Restoring the parameters of the dyonic solution, the result reads
\begin{equation}
\displaystyle
\widehat\alpha(b) = \frac{4M(1 - \ell)^{3/2}}{b}
 +\frac{3\pi}{4b^2}\left[
 5M^2(1 - \ell)^3 - Q^2 -\frac{(1-\ell)^2}{1-2\ell}p^2
 \right] +O(\epsilon^3).
\label{eq:dyonicdeflection}
\end{equation}
The corresponding coordinate angular excess is obtained from Eq.~\eqref{eq:angleconventions}. In particular, the Schwarzschild and dyonic Reissner--Nordstr\"om limits follow by setting $(\ell,Q,p)=(0,0,0)$ and $\ell=0$, respectively~\cite{Eiroa2002,Sereno2004}.


\subsection{Orbit, turning point, and Fermat calculations}
\label{subsec:checks}

In this subsection, an independent derivation follows from the null geodesic equations. The conserved quantities in Eq.~\eqref{eq:impact} give
\begin{equation}
\dot R^2 = \mathcal E^2 - \frac{A J^2}{R^2},\qquad
\left(\frac{\mathrm{d}U}{\mathrm{d}\varphi}\right)^2
 = \frac{1}{b^2} -U^2 + 2\mu U^3 - q^2U^4,\qquad
\frac{\mathrm{d}^2U}{\mathrm{d}\varphi^2}+U=3\mu U^2 - 2q^2 U^3,
\label{eq:Binet}
\end{equation}
where $U = 1/R$. Let $w = bU$, $h = \mu/b$, and $k = q^2/b^2$, and write $w = w_0 + hw_1 + h^2w_2 + kw_q + O(\epsilon^3)$. Fixing the incoming asymptote by $w(0)=0$ and $w'(0)=1$, we obtain
\begin{align}
w_0 = {}&\sin\varphi,\qquad w_1=(1 - \cos\varphi)^2,\nonumber\\
w_2 = {}&\frac{5}{16}\sin\varphi + 2\sin2\varphi
 -\frac{3}{16}\sin3\varphi-\frac{15}{4}\varphi\cos\varphi,\nonumber\\
w_q = {}& - \frac{9}{16}\sin\varphi-\frac{1}{16}\sin3\varphi
 + \frac34\varphi\cos\varphi.
\label{eq:orbitsolutions}
\end{align}
These functions solve $w_1''+w_1=3w_0^2$, $w_2''+w_2=6w_0w_1$, and $w_q''+w_q=-2w_0^3$. The outgoing root is located at $\varphi=\pi+\widehat\alpha$. Since $w_1(\pi)=4$, $w_1'(\pi)=0$, $w_2(\pi)=15\pi/4$, and $w_q(\pi)=-3\pi/4$, this root reproduces Eq.~\eqref{eq:deflection}. The first two terms of $U = w/b$ coincide with the corrected integration boundary in Eq.~\eqref{eq:GBorbit}, as we should naturraly expect.

The same angle can be calculated without solving the orbit perturbatively. If $R_0$ is the outer turning point, then $b^2=R_0^2/A(R_0)$ and
\begin{equation}
\widehat\alpha=2\int_{R_0}^{\infty}
\frac{b\,\mathrm{d}R}{R^2\sqrt{1-b^2A(R)/R^2}}-\pi
 = 2\int_0^1\frac{\mathrm{d}z}{\sqrt{1-z^2-2h_0(1-z^3) + k_0(1-z^4)}}-\pi,
\label{eq:exactangle}
\end{equation}
where $z = R_0/R$, $h_0 = \mu/R_0$, and $k_0=q^2/R_0^2$. Expanding the integrand with its turning point fixed gives
\begin{equation}
\widehat\alpha = \frac{4\mu}{R_0}
 +\left(\frac{15\pi}{4}-4\right)\frac{\mu^2}{R_0^2}
 -\frac{3\pi q^2}{4R_0^2} + O(\epsilon^3),\qquad
R_0=b-\mu+\frac{q^2 - 3\mu^2}{2b} + O(\epsilon^3 b).
\label{eq:closestapproach}
\end{equation}
Substitution again yields Eq.~\eqref{eq:deflection}. This distinction between closest approach and impact parameter is necessary when comparing weak--lensing coefficients~\cite{KeetonPetters2005,IyerPetters2007}. The integral in Eq.~\eqref{eq:exactangle}, with $z=\sin\chi$, also shows a regular numerical approach. Extending its expansion by one order, it gives
\begin{equation}
\widehat\alpha = \frac{4\mu}{b} 
  +\frac{\pi(15\mu^2-3q^2)}{4b^2} 
  +\frac{128\mu^3/3-16\mu q^2}{b^3} 
  +O(\epsilon^4),
\label{eq:cubiccheck}
\end{equation}
which estimates the leading truncation error of the second order result. It is important to mention that the cubic term is regarded as an orbit integral check; the optical curvature calculation above has been carried out through second order.

On the other hand, the Fermat's principle supplies an additional (and complementary) way to derive it directly in terms of an effective refractive index~\cite{PerlickReview2004,Sereno2004}. In this regard, let us introduce an isotropic radial coordinate $\rho$ through
\begin{equation}
R = \rho + \mu + \frac{\mu^2 - q^2}{4\rho}.
\label{eq:isotropic}
\end{equation}
On the equatorial covering plane, the optical metric becomes $\mathrm{d}T^2=n^2(\rho)(\mathrm{d}\rho^2+\rho^2\mathrm{d}\varphi^2)$, where
\begin{equation}
n(\rho) = 
\frac{\left[1+\mu/\rho+(\mu^2-q^2)/(4\rho^2)\right]^2}
 {1-(\mu^2-q^2)/(4\rho^2)}
= 1 + \frac{2\mu}{\rho}+\frac{7\mu^2-3q^2}{4\rho^2}
 +O(\epsilon^3).
\label{eq:index}
\end{equation}
Let $z$ run along the incoming ray and $x(z)$ denote its transverse position, so that $\rho^2=x^2+z^2$. Varying $\int n\sqrt{1+(\mathrm{d}x/\mathrm{d}z)^2}\,\mathrm{d}z$ gives
\begin{equation}
x'' = ( 1 + x'^2)\left(\partial_x\ln n-x'\partial_z\ln n\right),
\qquad
\ln n=\frac{2\mu}{\rho}-\frac{\mu^2+3q^2}{4\rho^2}+O(\epsilon^3).
\label{eq:Fermateq}
\end{equation}
For $x(-\infty) = b$ and $x'(-\infty)=0$, the first order displacement is
\begin{equation}
\delta x_1(z) = -\frac{2\mu}{b}
\left(\sqrt{b^2 + z^2} + z\right).
\label{eq:Fermatpath}
\end{equation}
At second order, this must be inserted into the leading transverse gradient, and the longitudinal term $-x'\partial_z \ln n$ must also be taken into account. Here, the two terms contribute $2\pi\mu^2/b^2$ each to the positive bending angle. Together with the explicit second order index term, they give
\begin{equation}
\widehat\alpha = -x'(+ \infty) + O(\epsilon^3)
= \frac{4\mu}{b} - \frac{\pi(\mu^2+3q^2)}{4b^2}
 +\frac{4\pi\mu^2}{b^2}+O(\epsilon^3),
\label{eq:Fermatresult}
\end{equation}
in agreement with the preceding methods. The trajectory correction is fundamentally required in both the curvature and refractive index descriptions.


\subsection{Finite distance bending and the exact lens map}
\label{subsec:finite}

When the source and observer are kept at finite radii, the deflection must include the local direction of the ray at each endpoint~\cite{Ishihara2016,Ishihara2017,OnoAsada2019}. Let their normalized radii be $R_S$ and $R_O$, with one radial turning point between them. Static orthonormal observers measure
\begin{equation}
\sin\Psi_i = \frac{b\sqrt{A(R_i)}}{R_i},\qquad i\in\{O,S\}.
\label{eq:localangle}
\end{equation}
For a ray travelling from the source to the observer, the source angle is obtuse and the observer angle is acute: $\Psi_S=\pi-\arcsin[b\sqrt{A_S}/R_S]$ and $\Psi_O=\arcsin[b\sqrt{A_O}/R_O]$. The finite distance bending in the unwrapped angular convention is 
\begin{align}
\widehat\alpha_{\mathrm{FD}}
= {}&\Psi_O - \Psi_S+\Delta\varphi \sum_{i=O,S}\left[
 \int_{R_0}^{R_i}\frac{b\,\mathrm{d}R}{R^2\sqrt{1-b^2A(R)/R^2}}
 +\arcsin\!\left(\frac{b\sqrt{A_i}}{R_i}\right)
 \right] - \pi.
\label{eq:finiteexact}
\end{align}
It vanishes for $\mu = q = 0$ on a fixed covering space part. This angle is a comparison of endpoint directions with the specified background.

For an explicit expansion, define $x_i = b/R_i$, $s_i = \sqrt{1-x_i^2}$, and $\mathcal H=\pi-\arcsin x_O - \arcsin x_S$. Subtracting the two asymptotic tails in Eq.~\eqref{eq:exactangle} and expanding the endpoint angles gives
\begin{align}
\widehat\alpha_{\mathrm{FD}}
= {}&\frac{2\mu}{b}(s_O+s_S)
 + \frac{\mu^2}{4b^2}\left[
 15\mathcal H + \sum_{i=O,S}\frac{x_i(15-7x_i^2)}{s_i}\right] - \frac{3q^2}{4b^2}\left[
 \mathcal H + \sum_{i=O,S} x_i s_i\right] + O(\epsilon^3).
\label{eq:finiteexpanded}
\end{align}
Here $x_i < 1$ is kept fixed in the weak--field expansion, away from the endpoint turning point limit $s_i = 0$. The expression is symmetric under $O\leftrightarrow S$ and tends to Eq.~\eqref{eq:deflection} as both radii tend to infinity. For distant endpoints, its leading correction is
\begin{equation}
\widehat\alpha_{\mathrm{FD}} - \widehat\alpha
= - \mu b\left(R_O^{-2} + R_S^{-2}\right)
 + \frac{(q^2 - \mu^2)b}{2}\left(R_O^{-3} + R_S^{-3}\right)
 + \cdots,
\label{eq:finitedifference}
\end{equation}
where the displayed terms are understood through the same second weak--field order. Dispersive media would alter both the optical metric and the relation between frequency and local direction~\cite{CrisnejoGallo2018,Ovgun2025,Ovgun2026}; no plasma contribution is included here.

The global lens equation follows most directly from the same geodesic integral, based on the Perlick's construction~\cite{Perlick2004}. Place the observer at original azimuth $\phi_O = 0$, and let $\Phi_S$ be the source azimuth. For a chosen orientation $\sigma = \pm 1$ and winding integer $n$, the lens map for a ray with one turning point is
\begin{equation}
\Phi_S + 2\pi n
 = \frac{\sigma}{\sqrt a}\sum_{i=O,S}
\int_{R_0}^{R_i}\frac{b\,\mathrm{d}R}{R^2\sqrt{1 - b^2 A(R)/R^2}},
\qquad
b = \frac{R_O|\sin\theta|}{\sqrt{A_O}},\qquad
\sigma = \operatorname{sgn}\theta.
\label{eq:exactlensmap}
\end{equation}
The observed angle $\theta$ is measured from the inward radial direction; the orientation of $\Phi_S$ is chosen to agree with $\sigma$. Eq.~\eqref{eq:exactlensmap} keeps the physical $2\pi$ identification of $\phi$ and is applicable without interpreting the conical constant as a local force. It also gives a direct numerical prescription for image positions at finite distance. The weak images considered below correspond to the two primary approaches.


\subsection{Lens equation and weak-field observables}
\label{subsec:observables}

In order to obtain closed expressions for the observables, consider an isolated lens with $R_O,R_S\gg b\gg\mu$, small observed and source angles, and $|\sqrt a-1|\ll1$. Additionally, let the source lie close to the original antipodal direction, $\Phi_S = \pi - \delta$. Its physical transverse displacement on the source sphere is $\sqrt a\,R_S\delta$ to first order in $\delta$. It is convenient to define
\begin{equation}
D = \frac{R_S}{R_O + R_S},\qquad
\beta = \frac{\sqrt a\,R_S}{R_O + R_S}\,\delta, \qquad
\theta_c = \pi D(1 - \sqrt a).
\label{eq:sourcegeometry}
\end{equation}
These quantities refer to the stated static source--observer construction. In particular, $R_S$ is a lens centered radius.

For each primary orientation, the total unwrapped angular change satisfies
\begin{equation}
\sqrt a\,(\pi - \sigma\delta)
 = \pi + \widehat\alpha(b)
 - b \left(\frac1{R_O} + \frac1{R_S}\right) + \cdots,
\qquad b\simeq R_O|\theta|.
\label{eq:thinconstruction}
\end{equation}
This relation is obtained by expanding the endpoint tails of Eq.~\eqref{eq:exactlensmap}. It gives the weak lens equation on the conical background, as shown as follows
\begin{equation}
\beta = \theta - \theta_c\operatorname{sgn}\theta
 -  D\widehat\alpha(R_O|\theta|)\operatorname{sgn}\theta
= \theta-\theta_c\operatorname{sgn}\theta
 - \frac{\theta_E^2}{\theta}
 - \frac{\mathcal B}{\theta|\theta|}+\cdots,
\label{eq:lensequation}
\end{equation}
where
\begin{equation}
\theta_E^2 = \frac{4\mu D}{R_O}, \qquad
\mathcal K_2 = \frac{3\pi}{4}(5\mu^2-q^2), \qquad
\mathcal B = \frac{D\mathcal K_2}{R_O^2}.
\label{eq:lenscoefficients}
\end{equation}
The coefficient $\theta_c$ has been derived using the source position on the cone. It is not $D\pi(a^{-1/2} - 1)$, which would follow from inserting the coordinate angular excess into an otherwise Euclidean lens equation. For $a = 1$, Eq.~\eqref{eq:lensequation} reduces to the usual small angle charged lens equation~\cite{VirbhadraEllis2000,Sereno2004,KeetonPetters2005}.

The approximation in Eq.~\eqref{eq:lensequation} keeps the second weak--field bending term and the leading conical geometry, as it should be. It is controlled, for instance, by the joint near alignment counting $\mu/b\sim|\theta|\sim|\beta|\sim|\sqrt a - 1|$, with $b/R_i$ of the same small order and fixed $q/\mu$. Cubic angular terms, the leading finite distance bending correction, and the difference between $b$ and $R_O|\theta|$ enter one order beyond the terms retained. One remark worthy to be mention is that, uutside this hierarchy, Eq.~\eqref{eq:exactlensmap}, with Eq.~\eqref{eq:localangle}, should be used directly.

To describe the image positions, let us define
\begin{equation}
x = \frac{\theta}{\theta_E}, \qquad y = \frac{\beta}{\theta_E}, \qquad
c_*  = \frac{\theta_c}{\theta_E}, \qquad
\eta = \frac{\mathcal B}{\theta_E^3}
= \frac{\mathcal K_2}{4\mu R_O\theta_E}.
\label{eq:dimensionlesslens}
\end{equation}
Eq.~\eqref{eq:lensequation} becomes $y = x -c_*\operatorname{sgn}x - x^{-1} - \eta/(x|x|)$. Holding $c_*$ fixed, the zeroth order image on side $\sigma$ and its first correction are
\begin{equation}
x_{\sigma,0} = \frac{y + \sigma c_*+\sigma\sqrt{(y+\sigma c_*)^2+4}}{2},
\qquad
x_\sigma = x_{\sigma,0}
 + \frac{\sigma\eta}{1+x_{\sigma,0}^2}+O(\eta^2).
\label{eq:imagepositions}
\end{equation}
Notice that these expressions apply only when both image impact parameters remain in the weak--field domain and $|\eta|/[|x_{\sigma,0}|(1+x_{\sigma,0}^2)]\ll1$. Multiplying the truncated lens equation by powers of $x$ can generate additional formal roots near the lens; in this way, such roots are not controlled weak images.

For exact alignment, $\beta = 0$, spherical symmetry gives an Einstein ring. Its radius is
\begin{equation}
\theta_{\mathrm{ring},0}
= \frac{\theta_c+\sqrt{\theta_c^2 + 4\theta_E^2}}{2},\qquad
\theta_{\mathrm{ring}}
= \theta_{\mathrm{ring},0}
 + \frac{\mathcal B}{\theta_{\mathrm{ring},0}^2 + \theta_E^2}
 + O(\eta^2\theta_E).
\label{eq:ring}
\end{equation}
When $\theta_c = 0$, this reduces to
\begin{equation}
\theta_{\mathrm{ring}} = \theta_E
 + \frac{\mathcal K_2}{8\mu R_O} + O(\eta^2\theta_E),\qquad
\theta_E = \left[\frac{4M(1-\ell)^{3/2}D}{R_O}\right]^{1/2}.
\label{eq:ringlocal}
\end{equation}
The case $\theta_c = 0$ describes either $a = 1$ or a separately calibrated local reference in which the global conical mapping has been removed. On the black hole, $q^2\leq\mu^2$ and $\mathcal K_2 > 0$. The second order correction increases the ring radius relative to its leading value, although increasing the charge at fixed $\mu$ reduces that correction. In addition, for the source geometry in Eq.~\eqref{eq:sourcegeometry}, positive $\ell$ gives $\theta_c < 0$ and decreases the leading ring radius at fixed $\mu,R_O,R_S$.

The signed magnification is determined by the Jacobian of the angular map. In the same small angle construction, its tangential and radial eigenvalues are
\begin{align}
\lambda_t = {}&\frac{\beta}{\theta}
= 1 - \frac{\theta_c}{|\theta|} - \frac{\theta_E^2}{\theta^2}
 - \frac{\mathcal B}{|\theta|^3},\nonumber\\
\lambda_r = {}&\frac{\mathrm{d}\beta}{\mathrm{d}\theta}
= 1 + \frac{\theta_E^2}{\theta^2} + \frac{2\mathcal B}{|\theta|^3},
\qquad
\mathfrak m_\sigma = 
\left.[\lambda_t\lambda_r]^{-1}\right|_{\theta = \theta_\sigma}.
\label{eq:magnification}
\end{align}
The absolute magnifications are $|\mathfrak m_\sigma|$. All expressions obtained from the truncated lens map are to be expanded through first order in $\eta$. For an unresolved source, the total magnification, flux ratio, image separation, and angular centroid are shwon as follows
\begin{align}
\mathfrak m_{\mathrm{tot}}={}&|\mathfrak m_+|+|\mathfrak m_-|,
\qquad \mathcal R = \frac{|\mathfrak m_+|}{|\mathfrak m_-|},
\qquad \Delta\theta = \theta_+-\theta_-,\nonumber\\
\theta_{\mathrm{cent}} = {}&
\frac{|\mathfrak m_+|\theta_++|\mathfrak m_-|\theta_-}
 {|\mathfrak m_+|+|\mathfrak m_-|},
\qquad \delta\theta_{\mathrm{cent}}  =  \theta_{\mathrm{cent}} - \beta.
\label{eq:observables}
\end{align}
Eqs~\eqref{eq:imagepositions} and \eqref{eq:magnification} give these quantities for a nonzero conical term. The point source magnification diverges on the tangential configuration, i.e., $\beta = 0$; on the other hand, a finite source requires averaging over its angular profile.

For comparison with the usual weak lensing expressions, set $c_* = 0$ and take $y > 0$. Thereby
\begin{align}
\Delta\theta = {}&\theta_E \left[\sqrt{y^2+4} + \eta\right]
 + O(\eta^2\theta_E),\nonumber\\
\mathfrak m_{\mathrm{tot}} = {}&\frac{y^2+2}{y\sqrt{y^2+4}} + O(\eta^2),\qquad
\delta\theta_{\mathrm{cent}}
=\frac{\theta_E y}{y^2+2} + O(\eta^2\theta_E).
\label{eq:localobservables}
\end{align}
The absence of a first order correction to the total magnification and centroid is the familiar cancellation established by Keeton and Petters~\cite{KeetonPetters2005,KeetonPetters2006}. One additional remark deserves being pointed out: it does not remove the corrections to the individual images or their separation, and should not be imposed on Eq.~\eqref{eq:observables} when $c_*\ne0$.

Furthermore, the arrival time difference supplies an additional observable. An exact coordinate travel time for each ray with one turning point is
\begin{equation}
\mathcal T(b) = \sum_{i=O,S}\int_{R_0}^{R_i}
\frac{\mathrm{d}R}{A(R)\sqrt{1 - b^2A(R)/R^2}}, \qquad
\Delta\tau_O = \sqrt{A_O}\,[\mathcal T(b_-) -\mathcal T(b_+)].
\label{eq:exacttime}
\end{equation}
The factor $\sqrt{A_O}$ converts the normalized coordinate time to the proper time of the static observer. At the order of the thin lens approximation, the corresponding Fermat potential is written as
\begin{equation}
\tau(\theta) = \frac{R_O}{D}\left[
\frac12(\theta-\beta)^2 - \theta_c|\theta|
 - \theta_E^2\ln\!\left(\frac{|\theta|}{\theta_{\mathrm{ref}}}\right)
 + \frac{\mathcal B}{|\theta|}\right],
\label{eq:timepotential}
\end{equation}
where $\theta_{\mathrm{ref}}$ changes only an unobservable additive constant. Now by differentiating Eq.~\eqref{eq:timepotential}, it reproduces to Eq.~\eqref{eq:lensequation}. In dimensionless form, $\tau = 4\mu\,\mathcal U + \mathrm{const}$, with
\begin{equation}
\mathcal U(x,y) = \frac12(x-y)^2 - c_*|x|-\ln|x| + \frac{\eta}{|x|}.
\label{eq:dimensionlesstime}
\end{equation}
Since the zeroth order images are stationary points of the leading arrival time function, their shifts do not contribute to the delay at first order in $\eta$. Then,
\begin{equation}
\Delta\tau = 4\mu\left[
\mathcal U_0(x_{-,0},y) - \mathcal U_0(x_{+,0},y) 
+ \eta\left(\frac1{|x_{-,0}|} - \frac1{|x_{+,0}|}\right)
\right] + O(\eta^2\mu),
\label{eq:delaygeneral}
\end{equation}
where $\mathcal U_{0}$ denotes Eq.~\eqref{eq:dimensionlesstime} at $\eta = 0$. For $c_* = 0$ and $y > 0$, this becomes
\begin{equation}
\Delta\tau = 4\mu\left[
\frac{y}{2}\sqrt{y^2 + 4}
+\ln\!\left(\frac{\sqrt{y^2 + 4} + y}{\sqrt{y^2 + 4} - y}\right)
+\eta y\right] + O(\eta^2\mu),
\label{eq:delaylocal}
\end{equation}
in agreement with the charged lens and post Newtonian formalisms~\cite{Sereno2004,KeetonPetters2005,KeetonPetters2006}. Observer normalization and cosmological redshift factors must be restored for the corresponding observational configuration; the present expressions refer to the isolated, distant static observer.

At fixed $M, Q, p$, the local bending expands as
\begin{align}
\widehat\alpha = {}&\frac{4M}{b}\left(1 - \frac32\ell\right)
+ \frac{15\pi M^2}{4b^2}(1 - 3\ell)
- \frac{3\pi(Q^2+p^2)}{4b^2} 
+ O(\ell^2,\epsilon^3).
\label{eq:linearresponse}
\end{align}
At fixed normalized mass $\mu$, however, the first local correction specific to the magnetic sector is
\begin{equation}
\widehat\alpha(\mu, Q, p, \ell)
 - \widehat\alpha_{\mathrm{RN}}(\mu,Q^2 + p^2)
 = - \frac{3\pi p^2}{4b^2}\frac{\ell^2}{1 - 2\ell}
 + O\!\left(\frac{\mu p^2}{b^3}\frac{\ell^2}{1 - 2\ell}\right).
\label{eq:magneticsignature}
\end{equation}
In this regard, the local metric lensing determines $\mu$ and $q^2$, while the global source--observer mapping can additionally probe $a$. It cannot separate $Q$ and $p$, or determine their signs, from the local orbit alone, as we should naturally expect. Also, the equatorial reduction in Eq.~\eqref{eq:equatorialRN} makes this degeneracy exact for the metric null trajectories.


\section{Astrophysical bounds and observational prospects}
\label{sec:bounds}

The angular scales of Sgr~A$^*$ and M87$^*$ provide a natural setting in which to examine the observational content of the preceding results. Stellar astrometry constrains the gravitational potential around Sgr~A$^*$, while horizon scale imaging probes the critical null trajectories of both objects~\cite{GRAVITY2020,GRAVITY2022,GRAVITY2024,EHTM872019,EHTSgrI2022,EHTSgrVI2022}. These measurements concern distinct observables. A weak Einstein ring is formed by a background source close to alignment, whereas the bright EHT ring is produced by emission near the compact object. Its relation to the critical curve requires an emission model and an instrumental calibration~\cite{Gralla2019,Psaltis2020,PerlickTsupko2022}. We maintain this distinction when translating the geometry into parameter bounds.

Throughout this section, the observed radiation is assumed to follow the metric null geodesics specified in Sec.~\ref{sec:weak}, and the compact object is described by the static solution. In this way, the EHT comparisons below are conditional tests of this probe prescription and spherical geometry.


\subsection{Mass calibration and bounds on weak images}

It is convenient to introduce the angular gravitational radius and the dimensionless squared charge,
\begin{equation}
\theta_g = \frac{\mu}{R_O},\qquad
\zeta = \frac{q^2}{\mu^2}
 = \frac{Q^2}{M^2(1 - \ell)^3}
  + \frac{p^2}{M^2(1 - \ell)(1 - 2\ell)},\qquad
0\leq\zeta\leq1.
\label{eq:observationalparameters}
\end{equation}
Notice that the final inequality follows from the existence of a horizon. For numerical estimates we write $\mu = G\mathcal M/c^2$, where $\mathcal M$ is the mass associated with the normalized potential. A mass inferred from observations is compared with $\mu$, together with the adopted distance calibration. The mass and distance estimates obtained within general relativity furnish useful reference values; an inference at nonzero $\ell$ must fit those quantities within the same geometry.

In terms of Eq.~\eqref{eq:observationalparameters}, the lens equation becomes
\begin{equation}
\beta = \theta-\pi D\left[1 - (1 - \ell)^{-1/2}\right]
 \operatorname{sgn}\theta 
- \frac{4D\theta_g}{\theta}
 - \frac{3\pi D\theta_g^2}{4\theta|\theta|}(5 - \zeta),
\label{eq:observationallens}
\end{equation}
to the order established in Sec.~\ref{subsec:observables}. It follows that
\begin{equation}
\theta_E^2 = 4D\theta_g, \qquad
\mathcal B = \frac{3\pi}{4}D\theta_g^2(5 -\zeta), \qquad
3\pi D\theta_g^2 \leq\mathcal B\leq\frac{15\pi}{4}D\theta_g^2.
\label{eq:Bbounds}
\end{equation}
Even before imposing observations, the black hole condition gives a finite interval for the second order image correction. Holding $\mu$, $D$, $\ell$, and $\beta$ fixed, Eq.~\eqref{eq:imagepositions} yields
\begin{equation}
\theta_\sigma(\zeta) - \theta_\sigma(0)
 = - \sigma\frac{3\pi}{16}
 \frac{\zeta\theta_g}{1+x_{\sigma,0}^2} 
+ O(\eta^2\theta_E).
\label{eq:individualimagebound}
\end{equation}
The charge moves both weak images towards the optical axis. The magnitude of either displacement is bounded by the coefficient of $\zeta$ in Eq.~\eqref{eq:individualimagebound}. This comparison uses the uncharged solution at the same conical geometry; it does not attribute a change ofthe calibrated mass to the charge.

For a vanishing or independently removed conical offset, the aligned image and the separation of the two weak images obey
\begin{align}
\theta_E + \frac{3\pi}{8}\theta_g
&\leq\theta_{\mathrm{ring}}
\leq\theta_E + \frac{15\pi}{32}\theta_g,
\nonumber\\
\theta_E\sqrt{y^2 + 4} + \frac{3\pi}{4}\theta_g
&\leq\Delta\theta
\leq\theta_E\sqrt{y^2 + 4} + \frac{15\pi}{16}\theta_g.
\label{eq:weakimageintervals}
\end{align}
These inequalities are understood through the retained order, with theoretical errors of order $\eta^2 \theta_E$ and the endpoint corrections specified previously. The lower and upper limits correspond to $\zeta = 1$ and $\zeta = 0$, respectively. For a generic antipodal configuration, the conical term must instead be retained in Eqs.~\eqref{eq:imagepositions} and \eqref{eq:ring}. The construction is consistent with the perturbative treatment of charged lenses and with proposals to measure stellar lensing around Sgr~A$^*$~\cite{Sereno2004,KeetonPetters2005,KeetonPetters2006,BozzaMancini2012}.

The two image positions also supply a useful relation that does not require the unlensed source angle. Writing $u = \theta_+ > 0$ and $v = - \theta_- > 0$, subtraction of their lens equations gives
\begin{equation}
2\theta_c = u + v -\theta_E^2 \left(\frac1u+\frac1v\right)
 - \mathcal B \left(\frac1{u^2} + \frac1{v^2}\right).
\label{eq:pairconstraint}
\end{equation}
Once $\mathcal B$ and $\theta_c$ are determined, the inferred parameters are
\begin{equation}
\zeta = 5 - \frac{4\mathcal B}{3\pi D\theta_g^2}, \qquad
\ell = 1 - \left(1 - \frac{\theta_c}{\pi D}\right)^{-2}.
\label{eq:parameterestimators}
\end{equation}
A single image pair constrains the combination in Eq.~\eqref{eq:pairconstraint}; in other words, it does not determine both parameters independently. Additional source, a measured flux ratio, or a differential arrival time can supply further information, provided the source trajectory and distance are fitted. In particular, the cancellation of the first correction to the unresolved centroid and total magnification at $\theta_c = 0$ prevents, however, those two quantities alone from measuring $\mathcal B$ at this order~\cite{KeetonPetters2006}.


\subsection{Angular scales and conditional sensitivities}

For Sgr~A$^*$ we adopt the reference values $\mathcal M = 4.297\times10^6M_\odot$ and $R_O = 8.277\,\mathrm{kpc}$ from the multiple orbit analysis of the GRAVITY Collaboration~\cite{GRAVITY2022}. For M87$^*$ we use $\mathcal M = 6.5\times10^9M_\odot$ and $R_O = 16.8\,\mathrm{Mpc}$ as an illustrative scale~\cite{EHTM872019}. The latter mass is derived from EHT modeling and is used only to evaluate weak lensing signals. Independent stellar dynamical information must be used instead~\cite{Gebhardt2011,Psaltis2020}.

Table~\ref{tab:weakscales} considers a hypothetical background source at $R_S = 1\,\mathrm{pc}$ in the geometry of Eq.~\eqref{eq:sourcegeometry}, with $\ell = 0$ as the reference configuration. For $R_S\ll R_O$, $\theta_E$ scales as $R_S^{1/2}$, while the leading charge corrections to the aligned ring and the image separation are independent of $R_S$.

\begin{table}[tb]
\centering
\caption{Weak lensing scales for a hypothetical source at $R_S = 1\,\mathrm{pc}$. The second order mass correction is measured relative to $\theta_E$ at $\zeta = 0$. The charge shifts compare $\zeta = 1$ with $\zeta = 0$ at fixed normalized mass and geometry.}
\label{tab:weakscales}
\begin{tabular}{lcc}
\toprule
Quantity & Sgr~A$^*$ & M87$^*$ \\
\midrule
$\theta_g$ & $5.124\,\mu\mathrm{as}$ & $3.819\,\mu\mathrm{as}$ \\
$\theta_E$ & $22.60\,\mathrm{mas}$ & $0.4331\,\mathrm{mas}$ \\
$R_O\theta_E/\mu$ & $4.410\times10^3$ & $1.134\times10^2$ \\
$15\pi\theta_g/32$ & $7.546\,\mu\mathrm{as}$ & $5.624\,\mu\mathrm{as}$ \\
$|\Delta\theta_{\mathrm{ring}}|_{\mathrm{charge,max}}$
 & $1.509\,\mu\mathrm{as}$ & $1.125\,\mu\mathrm{as}$ \\
$|\Delta(\Delta\theta)|_{\mathrm{charge,max}}$
 & $3.018\,\mu\mathrm{as}$ & $2.250\,\mu\mathrm{as}$ \\
\bottomrule
\end{tabular}
\end{table}

To express the sensitivity near the uncharged reference, suppose $|\theta_c|\ll\theta_E$ and define the ring residual
\begin{equation}
\mathcal E_\theta 
= \theta_{\mathrm{ring}} - \theta_E - \frac{15\pi}{32}\theta_g
 = \frac{\theta_c}{2} - \frac{3\pi}{32}\zeta\theta_g
\simeq - \frac{\pi D}{4}\ell - \frac{3\pi}{32}\zeta\theta_g.
\label{eq:ringresidual}
\end{equation}
Terms quadratic in $\theta_c/\theta_E$, mixed conical and second order corrections, and higher weak field orders have been omitted in this last expansion. Eq.~\eqref{eq:ringresidual} displays the degeneracy directly: a reduced ring radius can arise from positive $\ell$, positive $\zeta$, or both. If the residual is consistent with zero within an effective error $n\sigma_{\mathrm{eff}}$, then
\begin{equation}
\left|\frac{\pi D}{4}\ell + 
\frac{3\pi}{32}\theta_g\zeta\right|
\lesssim n\sigma_{\mathrm{eff}}.
\label{eq:ringband}
\end{equation}
This is a sensitivity band for a specified measurement, not a bound supplied by the existing EHT images. Here $n$ denotes the chosen error multiplier, and $\sigma_{\mathrm{eff}}$ includes the uncertainties and covariances of the ring radius, mass, observer distance, and source distance, together with the theoretical and astrophysical errors.

For example, if $\zeta$ and the geometry were independently known, the sensitivity to a small conical parameter would be
\begin{equation}
\sigma_\ell \simeq \frac{4\sigma_{\mathrm{eff}}}{\pi D}
 = \begin{cases}
5.11\times10^{-8}\,[\sigma_{\mathrm{eff}}/(1\,\mu\mathrm{as})],
&\text{Sgr~A$^*$},\\
1.04\times10^{-4}\,[\sigma_{\mathrm{eff}}/(1\,\mu\mathrm{as})],
&\text{M87$^*$},
\end{cases}
\label{eq:ellsensitivity}
\end{equation}
for the source configuration in Table~\ref{tab:weakscales}. In contrast, setting $\ell = 0$ gives a nominal charge limit
\begin{equation}
\zeta\lesssim\frac{32n\sigma_{\mathrm{eff}}}{3\pi\theta_g},
\nonumber
\end{equation}
which improves on the horizon condition only if its right hand side is smaller than unity. An effective error of $1\,\mu\mathrm{as}$ would be insufficient to exclude any part of the black hole charge range at the $n = 1.96$ consistency level using the ring radius alone. The smaller values of $\sigma_\ell$ in Eq.~\eqref{eq:ellsensitivity} are conditional forecasts; no ring with the stated source geometry has been used to obtain them.

The distance calibration is particularly restrictive. With $\boldsymbol\lambda = (\ln\mu,\ln R_O,\ln R_S)$ and covariance $\boldsymbol\Sigma$, the uncertainty in the leading Einstein angle is
\begin{equation}
\frac{\sigma_{\theta_E}^2}{\theta_E^2}
 = \boldsymbol w^{\mathsf T}\boldsymbol\Sigma\boldsymbol w,
\qquad
\boldsymbol w = \left(\frac12, - 1 + \frac D2,\frac{1 - D}{2}\right)^{\mathsf T}.
\label{eq:Einsteinerror}
\end{equation}
Keeping this contribution below $1\,\mu\mathrm{as}$ requires fractional accuracies in the predicted $\theta_E$ better than $4.42\times10^{-5}$ for the Sgr~A$^*$ settings and $2.31\times10^{-3}$ for M87$^*$. GRAVITY highlights the relevance of microarcsecond astrometry, but the performance of a specified observing mode cannot be assigned without qualification to a faint secondary image or an extended Einstein ring~\cite{GRAVITY2017,BozzaMancini2012,GRAVITY2024}. In addition, source blending, stellar motion, the host potential, and departures from spherical symmetry must enter an observational fit.

An identified variable background source would provide a further test through its image delay. At $\theta_c = 0$, restoring the unit of time in Eq.~\eqref{eq:delaylocal} gives the charge dependent part
\begin{equation}
\Delta t(\zeta) - \Delta t(0)
 = - \frac{3\pi}{4}\frac{\mu}{c}
 \frac{\theta_g}{\theta_E}\,\zeta y
 + O\!\left(\frac{\mu}{c}\eta^2\right).
\label{eq:delaychargeobservable}
\end{equation}
For $y = 1$, its magnitude at $\zeta = 1$ is approximately $11.3\,\mathrm{ms}$ for Sgr~A$^*$ and $665\,\mathrm{s}$ for M87$^*$ in Table~\ref{tab:weakscales}. These values describe two images of the same background event. It is important to mention that joint astrometry and timing would be valuable because the angular and temporal observables depend differently on the mass and source geometry~\cite{KeetonPetters2005,KeetonPetters2006}.


\subsection{About the consistency with horizon scale imaging}

The exact critical impact parameter supplies an additional constraint without extending the weak field expansion into the photon sphere region. Defining $x_{\mathrm{ph}} = R_{\mathrm{ph}}/\mu$, Eq.~\eqref{eq:critical} becomes
\begin{equation}
x_{\mathrm{ph}} = \frac{3+\sqrt{9 - 8\zeta}}{2},\qquad
h(\zeta) = \frac{b_{\mathrm{ph}}}{\mu}
 = \left(\frac{2x_{\mathrm{ph}}^3}{x_{\mathrm{ph}} - 1}\right)^{1/2}.
\label{eq:shadowfunction}
\end{equation}
For a static observer outside the photon sphere, the angular diameter of the critical curve reads
\begin{equation}
\Theta_{\mathrm{sh}} = 2\arcsin\!\left[
\frac{\mu}{R_O}\sqrt{A_O}\,h(\zeta)\right]
\simeq2\theta_g h(\zeta).
\label{eq:shadowangularsize}
\end{equation}
The last expression applies to the distant observer used here. The local angular projection has already been included; multiplying it by an additional factor of $\sqrt a$ would count the angular normalization twice. In this manner, at fixed $\theta_g$, the shadow depends on $\ell$, $Q$, and $p$ only through $\zeta$. As we can realize, this is the same local degeneracy encountered in the weak null trajectories.

The fractional deviation from the Schwarzschild diameter is
\begin{equation}
\delta_{\mathrm{sh}}(\zeta) = \frac{h(\zeta)}{3\sqrt3} - 1, 
\qquad
\frac{\mathrm{d}\delta_{\mathrm{sh}}}{\mathrm{d}\zeta}
 = - \frac{1+\delta_{\mathrm{sh}}}{x_{\mathrm{ph}}(x_{\mathrm{ph}} - 1)} < 0.
\label{eq:shadowdeviation}
\end{equation}
It decreases from zero at $\zeta = 0$ to $4/(3\sqrt3) - 1\simeq - 0.2302$ at extremality. Hence a lower observational limit on the calibrated shadow diameter gives an upper limit on $\zeta$. This use of an independently calibrated mass scale follows the horizon scale tests developed in Refs.~\cite{Psaltis2020,EHTSgrVI2022,Vagnozzi2023}.

For Sgr~A$^*$, the EHT fiducial analysis reports
\begin{equation}
\delta_{\mathrm{sh}} = -0.08^{+0.09}_{-0.09}
\quad\text{(VLTI mass prior)},\qquad
\delta_{\mathrm{sh}} = -0.04^{+0.09}_{-0.10}
\quad\text{(Keck mass prior)},
\label{eq:EHTSgrintervals}
\end{equation}
at the quoted $68\%$ credible level~\cite{EHTSgrVI2022}. For M87$^*$, the 2017 analysis with a stellar dynamical mass prior gives $\delta_{\mathrm{sh}} = -0.01\pm0.17$ at the same quoted level~\cite{EHTM872019}. Mapping the endpoints of these published intervals through Eq.~\eqref{eq:shadowdeviation} gives Table~\ref{tab:shadowbounds}.

\begin{table}[tb]
\centering
\caption{Conditional ranges obtained by mapping the published $68\%$ shadow size intervals into the static metric, with $0\leq\zeta\leq1$.}
\label{tab:shadowbounds}
\begin{tabular}{lccc}
\toprule
Observation and mass prior & $\delta_{\mathrm{sh}}$ interval
 & $\zeta_{\max}$ & $(q/\mu)_{\max}$ \\
\midrule
Sgr~A$^*$, VLTI & $[-0.17,\,0.01]$ & $0.8144$ & $0.9024$ \\
Sgr~A$^*$, Keck & $[-0.14,\,0.05]$ & $0.7011$ & $0.8373$ \\
M87$^*$, stellar dynamics & $[-0.18,\,0.16]$ & $0.8492$ & $0.9215$ \\
\bottomrule
\end{tabular}
\end{table}

These comparisons inherit the emission calibration and mass assumptions of the original analyses. Doubling the quoted errors as an approximate consistency check includes the entire interval $-0.2302\leq\delta_{\mathrm{sh}}\leq0$ in all three cases. It therefore gives no exclusion beyond the horizon condition. A $95\%$ credible bound on $\zeta$ would require the full likelihood, an explicit charge prior, and a treatment of the physical boundary; it cannot be inferred by relabeling Table~\ref{tab:shadowbounds}.

More recent M87$^*$ observations confirm the persistence of the emission ring~\cite{EHTM872024}. The analysis of the 2017, 2018, and 2021 years reported an average diameter of $43.9\pm0.6\,\mu\mathrm{as}$~\cite{EHTM872025}. This uncertainty characterizes the fitted bright ring. Updating the bounds requires the ring to shadow calibration and an independent mass to distance likelihood, including their systematic errors. Substituting this diameter directly into Eq.~\eqref{eq:shadowangularsize} would assign the emission structure the role of a geometric critical curve therefore.

In the original parameters, any chosen limit $\zeta\leq\zeta_{\max}$ reads
\begin{equation}
Q^2 + \frac{(1 - \ell)^2}{1 - 2\ell}\,p^2
\leq\zeta_{\max}\mu^2 
= \zeta_{\max}M^2(1 - \ell)^3.
\label{eq:chargeellipse}
\end{equation}
For fixed $\ell$ and $\mu$, this is an ellipse in the electric--magnetic charge plane. For example, at fixed $\mu,Q,p\ne0$, defining $W = (\zeta_{\max}\mu^2-Q^2)/p^2$, Eq.~\eqref{eq:chargeellipse} requires $W\geq1$ and
\begin{equation}
1 - W - \sqrt{W(W - 1)}\leq\ell
\leq1 - W + \sqrt{W(W - 1)},\qquad \ell < \frac12.
\label{eq:conditionalellcharge}
\end{equation}


\subsection{A complementary estimate from stellar precession}

The conical parameter can also be tested through timelike motion. This offers a useful complement to the local shadow scale because a bound orbit samples the global angular identification. Assume a neutral, minimally coupled star and write $U = 1/R$, with specific angular momentum $\mathcal L = R^2 \mathrm{d}\varphi/\mathrm{d}\tau$. The orbit equation on the covering plane reads
\begin{equation}
\frac{\mathrm{d}^2U}{\mathrm{d}\varphi^2} + U
 = \frac{\mu}{\mathcal L^2}
 - \frac{q^2}{\mathcal L^2}U + 3\mu U^2 - 2q^2U^3.
\label{eq:timelikeorbit}
\end{equation}
Let $P_\star$ denote the radial semilatus rectum, equal to the Newtonian semimajor axis times $1-e^2$ at leading order, and put $\epsilon_\star=\mu/P_\star\ll1$. Perturbing the Kepler ellipse gives an advance $\pi(6 - \zeta)\epsilon_\star$ in $\varphi$. Since the original azimuth has period $2\pi$, the advance relative to that identification is
\begin{align}
\Delta\phi_\star
& = 2\pi\left(\sqrt{1 - \ell} - 1\right)
 +\pi\sqrt{1 - \ell}\,(6 - \zeta)\epsilon_\star
 +O(\epsilon_\star^2),\nonumber\\
& = 6\pi\epsilon_\star - \pi\ell -\pi\zeta\epsilon_\star
 +O(\ell^2,\ell\epsilon_\star,\epsilon_\star^2).
\label{eq:precessioncone}
\end{align}
Positive $\ell$ and positive $\zeta$ both reduce the Schwarzschild advance. A comparison with the observed sky trajectory additionally requires the orbital projection and light propagation in this geometry. Equation~\eqref{eq:precessioncone} supplies the leading secular estimate for that comparison.

The GRAVITY observations established the Schwarzschild precession of S2 and improved the orbital constraints~\cite{GRAVITY2020,GRAVITY2022,GRAVITY2024}. For a direct estimate based on a single orbit, we use the 2024 S2--only result $f_{\mathrm{SP}} = 0.918\pm0.128$, obtained with the stated reference frame priors, where $f_{\mathrm{SP}}$ multiplies the Schwarzschild advance~\cite{GRAVITY2024}. The same work found a tighter value from several stars fitted with a common $f_{\mathrm{SP}}$.

Using the reference Schwarzschild advance $6\pi\epsilon_\star = 12.1$ arcmin per S2 orbit gives $\epsilon_\star\simeq1.8673\times10^{-4}$. To leading order,
\begin{equation}
f_{\mathrm{SP}} \simeq1 - \frac{\zeta}{6}
-\frac{\ell}{6\epsilon_\star},\qquad
\ell+\epsilon_\star\zeta
=6\epsilon_\star(1 - f_{\mathrm{SP}}).
\label{eq:precessionconstraint}
\end{equation}
Approximating the reported error by a Gaussian and retaining $1.96$ standard deviations gives the conditional band as follows
\begin{equation}
- 1.89 \times10^{-4}
\lesssim\ell + 1.8673\times10^{-4}\zeta
\lesssim3.73\times10^{-4}.
\label{eq:S2band}
\end{equation}
For vanishing charge this becomes $-1.89\times10^{-4}\lesssim\ell\lesssim3.73\times10^{-4}$. Allowing the entire black hole range $0 \leq \zeta \leq 1$ gives the projected interval $- 3.76 \times10^{-4}\lesssim\ell\lesssim 3.73 \times10^{-4}$. In contrast, setting $\ell = 0$ leaves the full charge range compatible with this approximate S2 band.

The complementarity is then explicit. The local shadow scale constrains $\zeta$, stellar precession constrains approximately $\ell + \epsilon_\star\zeta$, and a weak ring constrains the combination in Eq.~\eqref{eq:ringband}. A joint analysis should fit the underlying orbital and lensing data with shared nuisance parameters. In particular, the Sgr~A$^*$ EHT posterior already contains a stellar dynamical mass prior, so multiplying it by that same prior again would duplicate information.


\section{Strong gravitational lensing}
\label{sec:strong}

Photons whose impact parameter approaches $b_{\mathrm{ph}}$ from above spend an increasingly long interval near the unstable circular orbit before returning to the exterior region. Their angular path can possibly contain several complete revolutions, which produce a sequence of relativistic images on each side of the optical axis. We examine this regime through Tsukamoto's formulation of the strong--deflection expansion~\cite{Bozza2001,Bozza2002,Tsukamoto2017}, keeping the charge dependence exact and retaining the normalized variables and metric probe prescription of Sec.~\ref{sec:weak}. Related applications to Lorentz--violating black holes provide useful comparisons~\cite{AraBumblebee2026,PereiraKR2026}.

The equatorial reduction in Eq.~\eqref{eq:equatorialRN} allows the singular radial integral to be treated on the covering plane. The angular identification must nevertheless be restored when connecting the trajectories to a source and an observer. In particular, one revolution in the original azimuth corresponds to $2\pi\sqrt a$ in $\varphi$. This distinction affects the separation, relative brightness, and arrival times of successive images even when the normalized local orbit is unchanged, as we shall be seeing in what follows


\subsection{Expansion near the critical trajectory}

Starting from the exact scattering integral in Eq.~\eqref{eq:exactangle}, introduce Tsukamoto's radial variable
\begin{equation}
z = 1 - \frac{R_0}{R}, \qquad
I(R_0)\equiv\widehat\alpha(R_0) + \pi 
= 2 \int_0^1 \frac{\mathrm{d}z}{\sqrt{F(z,R_0)}},
\label{eq:SDLvariable}
\end{equation}
where the turning point condition gives
\begin{align}
F(z,R_0)
& = A(R_0) - (1-z)^2A\!\left(\frac{R_0}{1-z}\right)
 = c_1(R_0)z + c_2(R_0)z^2 + c_3(R_0)z^3 + c_4(R_0)z^4,
\label{eq:SDLpolynomial}
\end{align}
with
\begin{align}
c_1& = 2 - \frac{6\mu}{R_0} + \frac{4q^2}{R_0^2},
&c_2& = - 1 + \frac{6\mu}{R_0} - \frac{6q^2}{R_0^2},\nonumber\\
c_3& = - \frac{2\mu}{R_0} + \frac{4q^2}{R_0^2},
&c_4& = - \frac{q^2}{R_0^2}.
\label{eq:SDLc}
\end{align}
The polynomial is exact. Its first two terms isolate the singular behavior, while the remaining terms determine the finite contribution. The choice of $z$ is particularly useful for charged geometries because the regular integral at the photon sphere can remarkably be evaluated analytically~\cite{TsukamotoGong2017,Tsukamoto2017}.

Let a subscript $m$ denote evaluation at $R_{\mathrm{ph}}$, and define the dimensionless quantities
\begin{equation}
\nu_m = \frac{q^2}{R_{\mathrm{ph}}^2}, \qquad
\Delta_m = 1 - 2\nu_m, \qquad
\gamma_m = - \frac{2\mu}{R_{\mathrm{ph}}} + 4\nu_m.
\label{eq:SDLcriticaldata}
\end{equation}
The photon sphere identity $R_{\mathrm{ph}}^2 - 3\mu R_{\mathrm{ph}} + 2q^2 = 0$ implies $c_{1m} = 0$ and $c_{2m} = \Delta_m$. On the black hole configuration, $0\leq\nu_m\leq1/4$, so $1/2\leq\Delta_m\leq1$. In this manner, the integrand of Eq.~\eqref{eq:SDLvariable} behaves as $2/(\sqrt{\Delta_m}\,z)$ at criticality. The divergence is logarithmic throughout this part, including the extremal solution; also, the outer photon orbit does not become marginally unstable when the horizons coincide.

We separate the integral into its divergent and regular parts,
\begin{align}
I_D(R_0)& = 2 \int_0^1 
\frac{\mathrm{d}z}{\sqrt{c_1z+c_2z^2}}
 = \frac{4}{\sqrt{c_2}}
\ln\!\left(\frac{\sqrt{c_2} + \sqrt{c_1 + c_2}}{\sqrt{c_1}}\right),
\nonumber\\
I_R(R_0)& = 2 \int_0^1\left[
\frac{1}{\sqrt{F(z,R_0)}} - 
\frac{1}{\sqrt{c_1z + c_2z^2}}\right]\mathrm{d}z,
\qquad I = I_D + I_R.
\label{eq:SDLsplit}
\end{align}
The expression for $I_D$ applies in a neighborhood of the outer photon sphere, where $c_1,c_2>0$ for $R_0>R_{\mathrm{ph}}$. To express the divergence in terms of the impact parameter, we set
\begin{equation}
\varepsilon_R = \frac{R_0}{R_{\mathrm{ph}}} - 1, \qquad
\varepsilon_b = \frac{b}{b_{\mathrm{ph}}} - 1.
\label{eq:SDLsmallparameters}
\end{equation}
Expansion of $c_1$ and of the turning point relation gives
\begin{equation}
c_1 = 2 \Delta_m\varepsilon_R + O(\varepsilon_R^2), \qquad
\varepsilon_b = \frac{\Delta_m}{2A_m}\varepsilon_R^2
 + O(\varepsilon_R^3).
\label{eq:SDLturningexpansion}
\end{equation}
The absence of a linear term in $b(R_0)$ follows from its minimum at $R_{\mathrm{ph}}$. Substituting Eq.~\eqref{eq:SDLturningexpansion} into $I_D$ determines its logarithmic and constant terms, as shown below
\begin{equation}
I_D = - \frac{1}{\sqrt{\Delta_m}}\ln\varepsilon_b
 + \frac{1}{\sqrt{\Delta_m}}\ln\!\left(\frac{2\Delta_m}{A_m}\right) 
+o(1).
\label{eq:SDLdivergentlimit}
\end{equation}
The quadratic relation in Eq.~\eqref{eq:SDLturningexpansion} is essential: the logarithmic coefficient expressed in $\varepsilon_R$ is twice the coefficient expressed in $\varepsilon_b$.


\subsection{Analytic coefficients of the strong--deflection angle}

At the critical radius, the regular contribution reduces to
\begin{equation}
I_R ( R_{\mathrm{ph}} ) = 2 \int_0^1 \left[
\frac{1}{z\sqrt{\Delta_m + \gamma_m z - \nu_m z^2}}
 - \frac{1}{\sqrt{\Delta_m}\,z}\right]\mathrm{d}z.
\label{eq:SDLregularintegral}
\end{equation}
The two terms must be combined before taking the lower endpoint. Their difference has the finite limit $-\gamma_m/(2\Delta_m^{3/2})$ inside the square brackets. For an explicit integration, we define
\begin{equation}
H_m(z) = 2\sqrt{\Delta_m}
\sqrt{\Delta_m + \gamma_m z - \nu_m z^2}
 + 2\Delta_m + \gamma_m z.
\label{eq:SDLH}
\end{equation}
The primitive of the first term is $-\Delta_m^{-1/2}\ln[H_m(z)/z]$. Since $H_m(0) = 4\Delta_m$ and
$H_m(1) = 2[1-\mu/R_{\mathrm{ph}} + \sqrt{A_m\Delta_m}]$, evaluation at the endpoints yields
\begin{equation}
I_R ( R_{\mathrm{ph}} )
 = \frac{2}{\sqrt{\Delta_m}}
\ln\!\left[
\frac{2\Delta_m}
{1 - \mu/R_{\mathrm{ph}} + \sqrt{A_m\Delta_m}}\right].
\label{eq:SDLregularclosed}
\end{equation}
No expansion in $q/\mu$ has been used. Combining Eqs.~\eqref{eq:SDLdivergentlimit} and \eqref{eq:SDLregularclosed}, the deflection angle takes the form
\begin{equation}
\widehat\alpha(b)
 = - \bar a\ln\varepsilon_b + \bar b
 + O\!\left(\varepsilon_b|\ln\varepsilon_b|\right),
\label{eq:SDLangle}
\end{equation}
where
\begin{align}
\bar a & = \frac{1}{\sqrt{\Delta_m}}
 = \frac{R_{\mathrm{ph}}}{\sqrt{R_{\mathrm{ph}}^2-2q^2}},
\nonumber\\
\bar b& = \bar a\ln\!\left[
\frac{8\Delta_m^3}
{A_m\left( 1 - \mu/R_{\mathrm{ph}}+\sqrt{A_m\Delta_m}\right)^2}
\right] - \pi.
\label{eq:SDLcoefficients}
\end{align}
These are the local coefficients in the angular convention of Eq.~\eqref{eq:angleconventions}. Their analytic form agrees with the charged strong--deflection result after the normalized parameter identification~\cite{TsukamotoGong2017,Tsukamoto2017}. The logarithm in the remainder of Eq.~\eqref{eq:SDLangle} must be maintained; additionally, the residual is not generally of order $\varepsilon_b$ alone~\cite{IyerPetters2007,Tsukamoto2023}.

Two limits shows direct checks:
\begin{align}
q = 0:\quad&\bar a = 1,\qquad
\bar b = \ln\!\left[216(7-4\sqrt3)\right] - \pi,
\nonumber\\
q^2 = \mu^2: \quad&\bar a = \sqrt2, \qquad
\bar b = 2\sqrt2\ln\!\left[4( 2 - \sqrt2)\right] - \pi.
\label{eq:SDLlimits}
\end{align}
The first line concerns the normalized uncharged trajectory; the global spacetime is Schwarzschild only when $a = 1$ as well. If the previously defined coordinate excess $\widehat\alpha_{\phi}^{\mathrm{coord}}$ is used instead, its coefficients are
\begin{equation}
\bar a_\phi = \frac{\bar a}{\sqrt a}, \qquad
\bar b_\phi = \frac{\bar b+\pi}{\sqrt a} - \pi.
\label{eq:SDLcoordinatecoefficients}
\end{equation}
Notice that this conversion keeps the conical contribution in the constant term and avoids mixing the two angular conventions.

The coefficient $\bar a$ also admits a local dynamical interpretation. The circular orbit frequency in the covering angle and the radial Lyapunov exponent, both measured with respect to $T$, are
\begin{equation}
\Omega_{\varphi,m} = \frac{\sqrt{A_m}}{R_{\mathrm{ph}}}, \qquad
\lambda_m = \frac{\sqrt{A_m\Delta_m}}{R_{\mathrm{ph}}}, \qquad
\bar a = \frac{\Omega_{\varphi,m}}{\lambda_m}.
\label{eq:SDLinstability}
\end{equation}
The physical winding frequency is $\Omega_{\phi,m} = \Omega_{\varphi,m}/\sqrt a$, so $\bar a_\phi = \Omega_{\phi,m}/\lambda_m$. This is a geodesic relation of the type underlying the connection between strong lensing and photon orbit instability~\cite{Stefanov2010}.

Table~\ref{tab:SDLcoefficients} collects representative values of the local coefficients and leading image hierarchy.

\begin{table}[tb]
\centering
\caption{Strong--deflection coefficients and leading winding observables for $\ell=0$. The charge variable $\zeta$ is the squared normalized charge already defined in Eq.~\eqref{eq:observationalparameters}. The magnitude difference uses the geometric sum estimate in Eq.~\eqref{eq:SDLfluxratio}.}
\label{tab:SDLcoefficients}
\begin{tabular}{cccccc}
\toprule
$\zeta$ & $b_{\mathrm{ph}}/\mu$ & $\bar a$ & $\bar b$
 & $e_1$ & $\Delta m$ \\
\midrule
$0.0$   & $5.19615$ & $1.00000$ & $-0.40023$ & $1.2515\times10^{-3}$ & $6.820$ \\
$0.5$ & $4.70960$ & $1.08204$ & $-0.40066$ & $2.0765\times10^{-3}$ & $6.301$ \\
$1.0$   & $4.00000$ & $1.41421$ & $-0.73320$ & $7.0036\times10^{-3}$ & $4.811$ \\
\bottomrule
\end{tabular}
\end{table}


\subsection{Finite distance lens equation and relativistic images}

Finite endpoint radii alter the regular contribution but leave the logarithmic coefficient unchanged, provided $R_O,R_S > R_{\mathrm{ph}}$ remain fixed in the critical limit~\cite{BozzaScarpetta2007,Ishihara2017}. Let $\mathcal J_i$ denote the asymptotic angular tail of a critical ray beyond endpoint $i$. The same primitive used above gives
\begin{equation}
Z_i = 1 - \frac{R_{\mathrm{ph}}}{R_i}, \qquad
\mathcal J_i = \bar a\ln\!\left[
\frac{H_m(Z_i)}{Z_iH_m(1)}\right], \qquad i\in\{O,S\}.
\label{eq:SDLtails}
\end{equation}
In this manner, the total unwrapped excursion between the endpoints satisfies
\begin{equation}
\Delta\varphi(b) 
= - \bar a\ln\varepsilon_b + \bar b + \pi
 - \mathcal J_O - \mathcal J_S
 + O\!\left(\varepsilon_b|\ln\varepsilon_b|\right).
\label{eq:SDLfiniteexcursion}
\end{equation}
For the finite distance deflection defined in Eq.~\eqref{eq:finiteexact}, the corresponding constant is
\begin{equation}
\bar b_{\mathrm{FD}} = \bar b - \mathcal J_O - \mathcal J_S
 + \sum_{i=O,S}\arcsin\!\left(\frac{b_{\mathrm{ph}}\sqrt{A_i}}{R_i}\right).
\label{eq:SDLfiniteconstant}
\end{equation}
Both expressions recover their asymptotic counterparts as $R_O,R_S\rightarrow\infty$. The endpoint angles in Eq.~\eqref{eq:SDLfiniteconstant} belong to the finite distance bending convention; notice that the source lens map uses the excursion in Eq.~\eqref{eq:SDLfiniteexcursion}.

Consider now the source direction $\Phi_S = \pi - \delta$ introduced previously. Label the two image orientations by $\sigma = \pm1$ and let $n\geq1$ count complete revolutions in the original azimuth. The two parts of Eq.~\eqref{eq:exactlensmap} require that
\begin{equation}
\Delta\varphi_{n\sigma} 
= \sqrt a\left[(2n + 1)\pi - \sigma\delta\right].
\label{eq:SDLwinding}
\end{equation}
Substitution into Eq.~\eqref{eq:SDLfiniteexcursion} remarkably gives an explicit finite distance solution,
\begin{align}
\varepsilon_{n\sigma}^{\mathrm{FD}}
& = \exp\!\left[
\frac{\bar b + \pi - \mathcal J_O - \mathcal J_S
 - \sqrt a[(2n + 1)\pi - \sigma\delta]}{\bar a}\right],
\nonumber\\
b_{n\sigma}
& = b_{\mathrm{ph}}\left[1 + \varepsilon_{n\sigma}^{\mathrm{FD}}
+O\!\left((\varepsilon_{n\sigma}^{\mathrm{FD}})^2
|\ln\varepsilon_{n\sigma}^{\mathrm{FD}}|\right)\right].
\label{eq:SDLfiniteimages}
\end{align}
The observed image follows from Eq.~\eqref{eq:localangle}, with its sign fixed by $\sigma$. Eqs.~\eqref{eq:SDLtails}--\eqref{eq:SDLfiniteimages} remain applicable without replacing the endpoint geometry by angular diameter distances.

For compact expressions for the usual observables, specialize to $R_O,R_S\gg b_{\mathrm{ph}}$ and a nearly aligned source. We retain the definitions of $D$ and $\beta$ in Eq.~\eqref{eq:sourcegeometry}. Subsequently, Eby expanding the endpoint tails as $b/R_i$ gives the lens equation
\begin{equation}
\beta = \theta - \sigma D\left[
\widehat\alpha(R_O|\theta|) - \alpha_n^{(a)}\right], \qquad
\alpha_n^{(a)} = \pi\left[(2n + 1)\sqrt a - 1\right].
\label{eq:SDLlensequation}
\end{equation}
For $a = 1$, the usual subtraction $2\pi n$ is recovered~\cite{VirbhadraEllis2000,Bozza2002}. For $a \ne 1$, Eq.~\eqref{eq:SDLlensequation} follows from the physical winding condition. The relation $b\simeq R_O|\theta|$ is used only in this distant observer approximation; Eq.~\eqref{eq:localangle} supplies its finite radius correction therefore.

The accumulation angle is half the critical curve diameter in Eq.~\eqref{eq:shadowangularsize}, $\theta_\infty = \Theta_{\mathrm{sh}}/2$, and tends to $b_{\mathrm{ph}}/R_O$ for a distant observer. Let us introduce
\begin{equation}
e_n = \exp\!\left(\frac{\bar b - \alpha_n^{(a)}}{\bar a}\right), \qquad
\theta_{n\sigma}^{(0)} = \sigma\theta_\infty(1 + e_n),\qquad
\lambda_n = \frac{\theta_\infty e_n}{\bar a D}.
\label{eq:SDLimageparameters}
\end{equation}
At $\theta_{n\sigma}^{(0)}$ the leading logarithmic deflection equals $\alpha_n^{(a)}$. Its first variation is
\begin{equation}
\widehat\alpha - \alpha_n^{(a)} 
\simeq - \frac{\sigma\bar a}{\theta_\infty e_n}
\left(\theta - \theta_{n\sigma}^{(0)}\right).
\label{eq:SDLlinearangle}
\end{equation}
Insertion into Eq.~\eqref{eq:SDLlensequation} it yields
\begin{equation}
\theta_{n\sigma} 
\simeq \frac{\theta_{n\sigma}^{(0)} + \lambda_n\beta}{1 + \lambda_n}
 = \theta_{n\sigma}^{(0)}
 + \frac{\lambda_n}{1 + \lambda_n}
\left(\beta - \theta_{n\sigma}^{(0)}\right).
\label{eq:SDLpositions}
\end{equation}
Keeping $1 + \lambda_n$ records the exact solution of the linearized lens equation. Besides $e_n\ll1$, the linearization requires $|\beta-\theta_{n\sigma}^{(0)}|/(\bar a D)\ll1$. At exact alignment the two orientations form the $n$--th relativistic Einstein ring, whose angular radius is
\begin{equation}
\theta_n^{\mathrm{E}} \simeq 
\frac{\theta_\infty(1 + e_n)}{1 + \lambda_n}.
\label{eq:SDLrings}
\end{equation}
It is worth pointing out that these rings accumulate at $\theta_\infty$ and are distinct from the weak Einstein ring discussed earlier.


\subsection{Magnifications, angular separation, and flux ratios}

The signed magnification follows from the angular Jacobian. Applying the definition already used in Sec.~\ref{subsec:observables} to Eq.~\eqref{eq:SDLpositions} gives
\begin{equation}
\mathfrak m_{n\sigma}
\simeq\frac{\theta_{n\sigma}}{\beta}
\frac{\lambda_n}{1+\lambda_n},\qquad
|\mathfrak m_{n\sigma}|
\simeq\frac{\theta_\infty^2}{\bar a D|\beta|}
e_n(1 + e_n),
\label{eq:SDLmagnification}
\end{equation}
where the second expression keeps the leading distant source contribution. For $\beta > 0$, the $\sigma = + 1$ image has positive parity and the $\sigma = - 1$ image has negative parity. Their absolute magnifications agree at leading order, while Eq.~\eqref{eq:SDLpositions} shows the first source position asymmetry. As in the weak case, notice that the divergence at $\beta = 0$ is regularized by the finite angular extent of the source.

The image sequence is encoded by the factor
\begin{equation}
\varrho = \exp\!\left( - \frac{2\pi\sqrt a}{\bar a}\right), \qquad
e_n = e_1 \varrho^{n-1}.
\label{eq:SDLhierarchy}
\end{equation}
At leading order, both the offsets from the critical curve and the fluxes decrease by $\varrho$ with each additional revolution. In an equivalent manner, the exponent is the radial instability accumulated during one physical orbit, since $2\pi\sqrt a/\bar a=\lambda_m(2\pi\sqrt a\,b_{\mathrm{ph}})$.

Now, let us suppose that the first relativistic image can be separated from the unresolved images with $n \geq 2$. On either side of the lens, define its angular separation from the accumulation point and its flux ratio by
\begin{equation}
s\equiv|\theta_{1\sigma}| - \theta_\infty
\simeq\theta_\infty e_1, \qquad
\mathcal R_{\mathrm{rel}} \equiv
\frac{|\mathfrak m_{1\sigma}|}
{\displaystyle\sum_{n=2}^{\infty}|\mathfrak m_{n\sigma}|}.
\label{eq:SDLobservables}
\end{equation}
The approximation for $s$ neglects the small source position correction in Eq.~\eqref{eq:SDLpositions}. Summing the leading geometric flux sequence gives
\begin{equation}
\mathcal R_{\mathrm{rel}} \simeq
\frac{1 - \varrho}{\varrho}
 = \exp\!\left(\frac{2\pi\sqrt a}{\bar a}\right) - 1, \qquad
\Delta m = 2.5 \log_{10}\mathcal R_{\mathrm{rel}}.
\label{eq:SDLfluxratio}
\end{equation}
The frequently used expression $\mathcal R_{\mathrm{rel}} \simeq \exp(2\pi\sqrt a/\bar a)$ follows when $\varrho\ll1$~\cite{Bozza2002}. Retaining the minus one performs the geometric sum before that additional approximation; corrections from the omitted bending terms and from the source geometry remain. The same leading ratio applies if each unresolved component contains the corresponding pair of images.

The inverse relations are
\begin{equation}
\bar a_\phi \simeq \frac{2\pi}{\ln(1 + \mathcal R_{\mathrm{rel}})}, \qquad
\bar b_\phi \simeq 2\pi + \bar a_\phi
\ln\!\left(\frac{s}{\theta_\infty}\right).
\label{eq:SDLinverse}
\end{equation}
Thereby, the usual strong lensing observables determine the coefficients associated with physical windings. Recovering the local coefficients $\bar a,\bar b$ additionally requires $a$. If the normalized mass to distance ratio is independently calibrated, the critical scale constrains $\zeta$ through the function $h(\zeta)$ already obtained in Sec.~\ref{sec:bounds}. The predicted $\bar a(\zeta)$ then allows a conditional reconstruction,
\begin{equation}
\sqrt a \simeq \frac{\bar a(\zeta)}{2\pi}
\ln(1 + \mathcal R_{\mathrm{rel}}), \qquad
\ell\simeq1 - 
\left[\frac{2\pi}
{\bar a(\zeta)\ln(1 + \mathcal R_{\mathrm{rel}})}\right]^2.
\label{eq:SDLellreconstruction}
\end{equation}


\subsection{Differential arrival times}

The exact travel time has already been given in Eq.~\eqref{eq:exacttime}. For fixed endpoint radii, Fermat's principle implies the useful identity
\begin{equation}
\frac{\mathrm{d}\mathcal T}{\mathrm{d}b}
 = b \,\frac{\mathrm{d}\Delta\varphi}{\mathrm{d}b}.
\label{eq:SDLtimeidentity}
\end{equation}
It can also be obtained by differentiating the two radial integrals together: their moving turning point contributions cancel because $\mathrm{d}T/\mathrm{d}\varphi = b$ at $R_0$. Now, let us combining Eq.~\eqref{eq:SDLtimeidentity} with the near critical excursion gives
\begin{equation}
\mathcal T(b)
 = b_{\mathrm{ph}}\Delta\varphi(b) + \mathcal T_{\mathrm{reg}}
 - \bar a b_{\mathrm{ph}}\varepsilon_b
 + O\!\left(b_{\mathrm{ph}}\varepsilon_b^2|\ln\varepsilon_b|\right),
\label{eq:SDLtimeexpansion}
\end{equation}
where $\mathcal T_{\mathrm{reg}}$ is finite and common to rays with the same endpoints. In particular, the coefficient of $-\ln\varepsilon_b$ in the travel time is $b_{\mathrm{ph}}\bar a$. The regular constant is unnecessary for differential timing as hilighted in~\cite{BozzaMancini2004}.

For two images on the same side with $n > k$, their angular excursions differ by exactly $2\pi\sqrt a(n - k)$. Restoring the unit of time and the observer's clock normalization gives
\begin{align}
\Delta\tau^{\mathrm{same}}_{n,k}
 = {}&\frac{\sqrt{A_O}\, b_{\mathrm{ph}}}{c}
\left[2\pi\sqrt a(n - k)
 + \bar a(\varepsilon_{k\sigma} - \varepsilon_{n\sigma})\right]
\nonumber\\
& + O\!\left(\frac{\sqrt{A_O}\,b_{\mathrm{ph}}}{c}
\left[\varepsilon_{k\sigma}^2|\ln\varepsilon_{k\sigma}|
 + \varepsilon_{n\sigma}^2|\ln\varepsilon_{n\sigma}|\right]\right).
\label{eq:SDLtimedelay}
\end{align}
Here $\varepsilon_{j\sigma} = b_{j\sigma}/b_{\mathrm{ph}} - 1$ denotes the actual image displacement; Eq.~\eqref{eq:SDLfiniteimages} supplies its leading approximation with an error of the displayed order. On the other hand, the leading term is obtained by replacing $2\pi(n - k)$ with $2\pi(n - k) - (\sigma - \sigma')\delta$ and using the corresponding impact parameters.

Combining the leading delay of adjacent images with the critical angular radius gives a distance relation,
\begin{equation}
\Delta\tau_{n + 1,n} \simeq
\frac{2\pi\sqrt a\,R_O}{c}\sin\theta_\infty, 
\qquad 
R_O \simeq \frac{c\,\Delta\tau_{n + 1, n}}
{2\pi\sqrt a\sin\theta_\infty}.
\label{eq:SDLdistance}
\end{equation}
The factors of $\sqrt{A_O}$ cancel between the angular projection and the measured time. In the distant observer limit, $\sin\theta_\infty$ can be replaced by $\theta_\infty$. This extends the usual strong lensing distance estimator to the specified conical geometry~\cite{BozzaMancini2004}.

The validity of the logarithmic approximation can be assessed directly using Eq.~\eqref{eq:SDLvariable}, with $z = t^2$ to regularize the turning point. For $\varepsilon_b = 10^{-5}$, the absolute difference between the exact bending and Eq.~\eqref{eq:SDLangle} is $4.40\times10^{-5}$, $4.45\times10^{-5}$, and $7.86\times10^{-5}$ radians for $\zeta = 0,0.5,1$, respectively. At $\varepsilon_b = 10^{-7}$ these differences decrease to $5.68\times10^{-7}$, $5.71\times10^{-7}$, and $1.03\times10^{-6}$ radians. In addition, finite distance image positions and differential times can likewise be checked with the exact map and travel time integral already available in the manuscript. In this regard, such a check is especially useful for the first image, whose displacement from criticality is largest~\cite{Tsukamoto2023}.

\begin{figure}[tbp]
  \centering
    \includegraphics[scale=0.44]{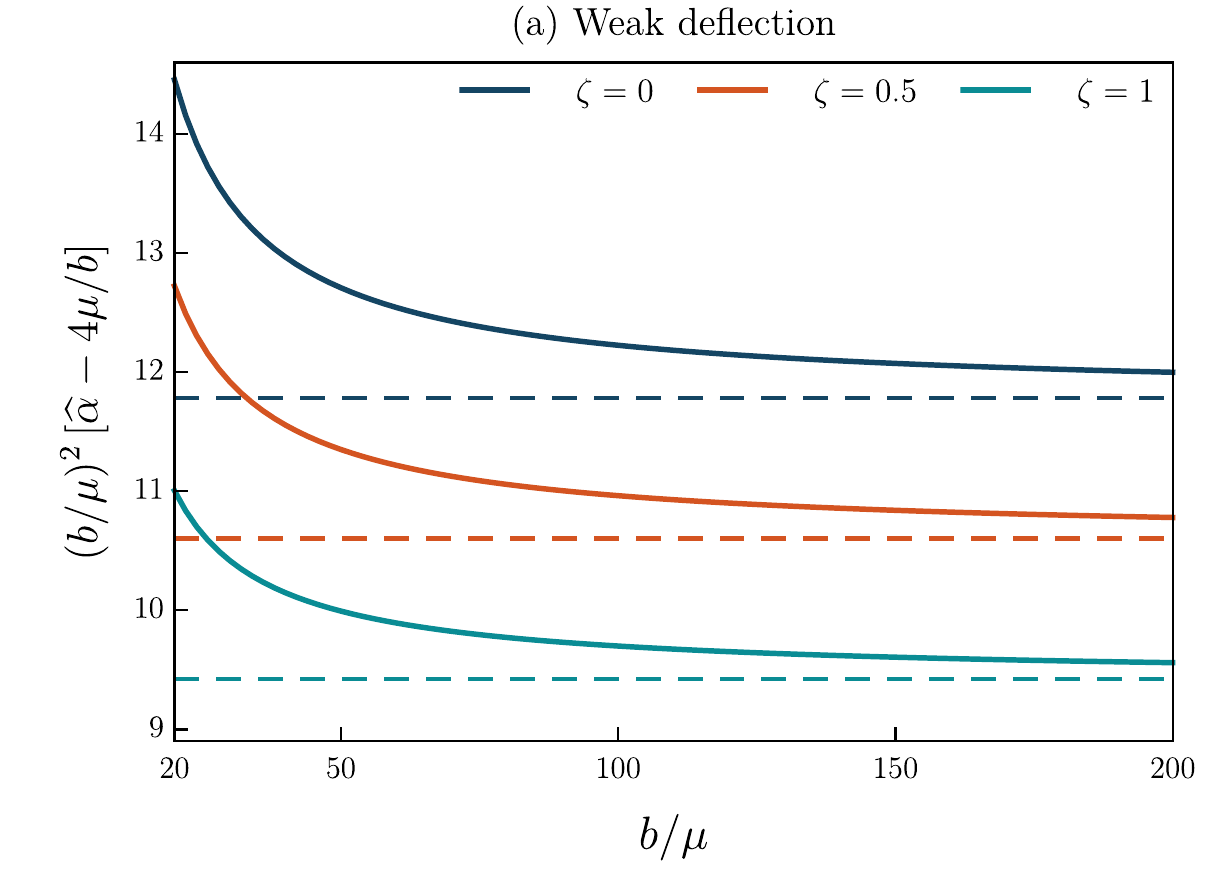}
  \includegraphics[scale=0.44]{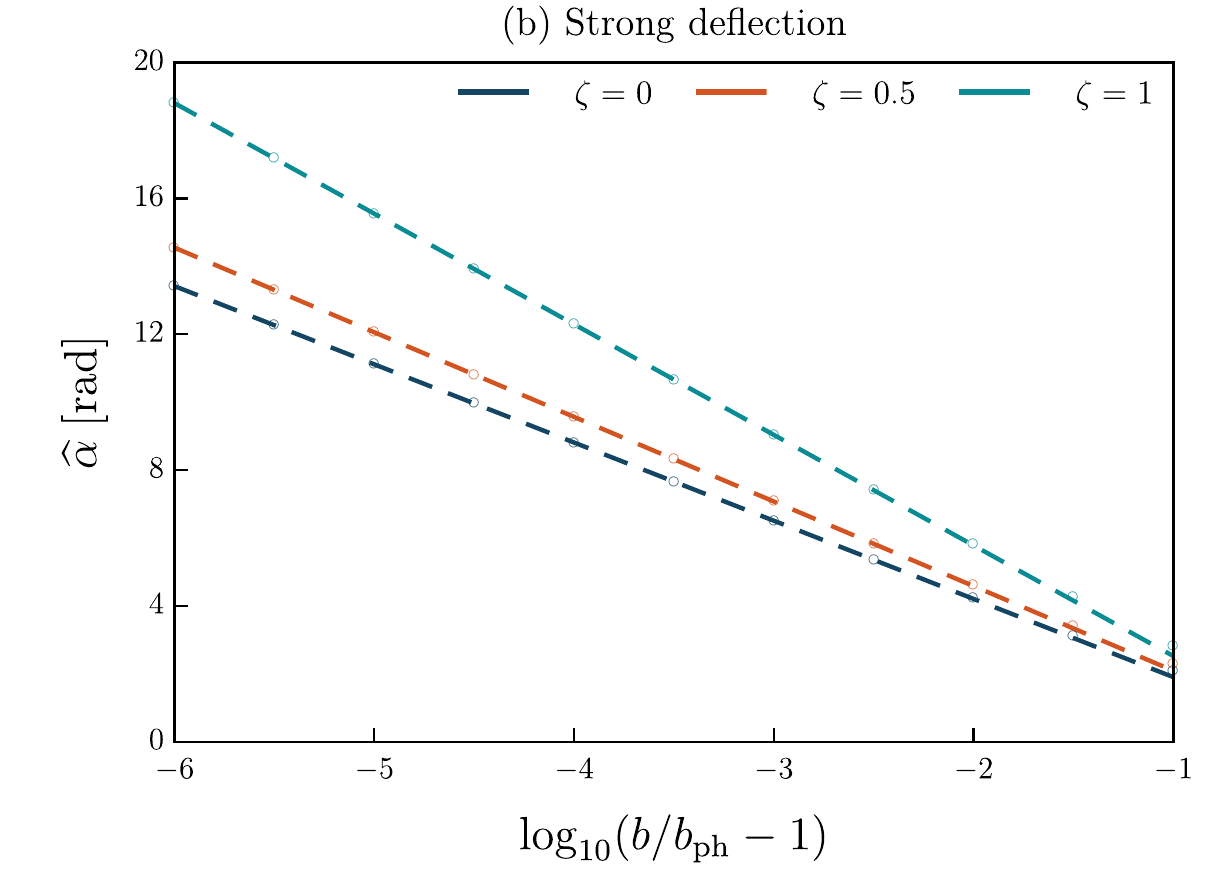}
  \caption{Weak and strong light bending for $\zeta=0,0.5,1$ at fixed $\mu$. (a) Reduced angle $(b/\mu)^2[\widehat\alpha - 4\mu/b]$:   solid curves give the exact integral and dashed lines its second-order limit $3\pi(5-\zeta)/4$. (b) Near critical bending: circles give the exact integral and dashed curves the logarithmic approximation. Each curve uses its own $b_{\mathrm{ph}}(\zeta)$.}
  \label{fig:kr-bending}
\end{figure}

In Fig.~\ref{fig:kr-bending}, the exact scattering integral is compared with the weak-- and strong--deflection expansions at fixed normalized mass. The reduced weak deflection signal approaches $3\pi(5 - \zeta)/4$, making the charge suppression explicit. Near criticality, the logarithmic coefficient increases with $\zeta$; this comparison holds $b/b_{\mathrm{ph}} - 1$ fixed, instead of $b/\mu$.

\section{Observational bounds in the strong deflection regime}
\label{sec:strongbounds}

The relativistic image sequence shows both the critical orbit and the angular advance accumulated during successive windings. Its observables contain information beyond the critical angular scale alone. We now translate the results of Sec.~\ref{sec:strong} into allowed ranges and parameter constraints, retaining the mass calibration of Sec.~\ref{sec:bounds}. The published horizon scale measurements supply conditional restrictions on the critical curve~\cite{EHTM872019,EHTSgrVI2022,Psaltis2020}; resolving the image hierarchy or identifying delayed copies of the same emission event would supply additional, independent observables~\cite{Bozza2002,BozzaMancini2004,Johnson2020,Hadar2021}. The latter possibilities are treated here as prospective measurements.


\subsection{Allowed ranges and astrophysical scales}

The coefficients in Eq.~\eqref{eq:SDLcoefficients} obey $1 \leq \bar a\leq\sqrt2$ on the black hole. Together with the critical scale already obtained in Eq.~\eqref{eq:shadowfunction}, this gives
\begin{equation}
4 \theta_g\leq\theta_\infty\leq3\sqrt3\,\theta_g,
 \qquad 
\exp(\sqrt2\pi\sqrt a) - 1
\leq\mathcal R_{\mathrm{rel}}
\leq\exp(2\pi\sqrt a) - 1,
\label{eq:SBbranchranges}
\end{equation}
at the distant observer and leading strong--deflection orders used for these observables. The inequalities delimit the predictions at fixed normalized mass, distance, and $a$. The image orders used to estimate the flux contrast must satisfy the strong--deflection condition specified in Sec.~\ref{sec:strong}.

For the same reason, the leading delay between adjacent images on one side satisfies
\begin{equation}
\frac{8\pi\mu\sqrt a}{c}
 \leq \Delta\tau_{n + 1,n}
 \leq \frac{6\pi\sqrt3\,\mu\sqrt a}{c}.
\label{eq:SBdelayrange}
\end{equation}
Eq.~\eqref{eq:SDLtimedelay} gives the exponentially small correction and the finite radius clock factor when either is required. The endpoints in Eqs.~\eqref{eq:SBbranchranges} and \eqref{eq:SBdelayrange} correspond to the extremal and uncharged normalized geometries.

Table~\ref{tab:SDLscales} gives the angular and temporal scales for the Sgr~A$^*$ and M87$^*$ benchmarks introduced earlier. The separation $s$ is much smaller than the critical angular radius, whereas an additional complete winding produces delays of minutes for Sgr~A$^*$ and days for M87$^*$. Notice that the critical curve, the relativistic images of a background source, and the bright accretion flow ring are different quantities~\cite{Gralla2019,PerlickTsupko2022}.

\begin{table}[tb]
\centering
\caption{Illustrative strong lensing scales at $\ell = 0$, using the mass calibration of Sec.~\ref{sec:bounds}. The entries are leading theoretical predictions for near alignment and distant endpoints.}
\label{tab:SDLscales}
\begin{tabular}{lcccc}
\toprule
Object & $\zeta$ & $\theta_\infty\ (\mu\mathrm{as})$
 & $s\ (\mu\mathrm{as})$ & $\Delta\tau_{n+1,n}$ \\
\midrule
Sgr~A$^*$ & $0$ & $26.627$ & $0.0333$ & $11.517\,\mathrm{min}$ \\
 & $0.5$ & $24.134$ & $0.0501$ & $10.438\,\mathrm{min}$ \\
 & $1$ & $20.497$ & $0.1436$ & $8.866\,\mathrm{min}$ \\
\midrule
M87$^*$ & $0$ & $19.844$ & $0.0248$ & $12.098\,\mathrm{days}$ \\
 & $0.5$ & $17.986$ & $0.0373$ & $10.965\,\mathrm{days}$ \\
 & $1$ & $15.276$ & $0.1070$ & $9.313\,\mathrm{days}$ \\
\bottomrule
\end{tabular}
\end{table}

At fixed $\mu$, $R_O$, and $\zeta$, the dependence on the conical parameter is particularly simple:
\begin{equation}
\frac{s(\ell)}{s(0)} \simeq 
\exp\!\left[-\frac{3\pi}{\bar a}(\sqrt a-1)\right],\qquad
\frac{\Delta\tau_{n + 1,n}(\ell)}
{\Delta\tau_{n + 1,n}(0)}\simeq\sqrt a.
\label{eq:SDLconicalresponse}
\end{equation}
The critical angle is unchanged under this comparison. Positive $\ell$ decreases the separation from the critical curve, increases the leading flux contrast, and lengthens the winding delay. A remark is needed to be pointed out at this point: holding $M$ instead of $\mu$ would also change the local angular scale, so a parameter bound must specify which mass calibration is used.


\subsection{Conditional bounds from the calibrated shadow size}

The charge intervals in Table~\ref{tab:shadowbounds} can be propagated directly through the strong--deflection coefficients. For any adopted limit $0\leq\zeta\leq\zeta_{\max}$, monotonicity of $h(\zeta)$ and $\bar a(\zeta)$ gives
\begin{align}
\theta_g h(\zeta_{\max})&\leq\theta_\infty
\leq3\sqrt3\,\theta_g,
&1&\leq\bar a\leq\bar a(\zeta_{\max}),\nonumber\\
\exp\!\left[\frac{2\pi\sqrt a}{\bar a(\zeta_{\max})}\right] - 1
&\leq\mathcal R_{\mathrm{rel}}
\leq\exp(2\pi\sqrt a) - 1.
\label{eq:SBshadowpropagation}
\end{align}
These are restrictions inherited from the same shadow size comparison. Their charge content remains the combination in Eq.~\eqref{eq:chargeellipse}; neither the electric and magnetic charges nor their signs are separated by metric lensing, as already mentioined in this paper.

For $\ell = 0$, on the other hand, the first image displacement $e_1$ increases over $0\leq\zeta\leq1$, and the corresponding intervals are listed in Table~\ref{tab:SBconditional}. The lower endpoint of $s/\theta_\infty$ is $1.2515\times10^{-3}$ in every row, while the upper endpoint of $\Delta m$ is $6.820$. In this case, even a shadow compatible charged solution can appreciably change the relative separation and brightness of the higher order images. The table uses the published $68\%$ shadow size intervals only as conditional ranges; its entries are not newly inferred $68\%$ credible limits on the strong lensing observables~\cite{EHTM872019,EHTSgrVI2022,Vagnozzi2023}.

\begin{table}[tb]
\centering
\caption{Propagation of the charge limits in Table~\ref{tab:shadowbounds} into leading strong lensing predictions at $\ell = 0$. The two Sgr~A$^*$ rows use alternative mass priors on the same imaging data. The last column is the common lower endpoint of $\theta_\infty/(3\sqrt3\,\theta_g)$ and of the adjacent image delay divided by $6\pi\sqrt3\,\mu/c$; their upper endpoint is unity.}
\label{tab:SBconditional}
\begin{tabular}{lccccc}
\toprule
Calibration & $\zeta_{\max}$ & $\bar a_{\max}$
 & $(s/\theta_\infty)_{\max}$ & $(\Delta m)_{\min}$ & Scale ratio \\
\midrule
Sgr~A$^*$, VLTI & $0.8144$ & $1.2048$ & $3.6976\times10^{-3}$ & $5.656$ & $0.83$ \\
Sgr~A$^*$, Keck & $0.7011$ & $1.1465$ & $2.8783\times10^{-3}$ & $5.945$ & $0.86$ \\
M87$^*$, stellar & $0.8492$ & $1.2287$ & $4.0543\times10^{-3}$ & $5.545$ & $0.82$ \\
Horizon only & $1$ & $1.4142$ & $7.0036\times10^{-3}$ & $4.811$ & $0.7698$ \\
\bottomrule
\end{tabular}
\end{table}

The emission remains relevant to these bounds. In charged geometries, the response of a bright lensing ring need not follow that of the critical curve, and changing the emission prescription can relax a charge constraint~\cite{TsukamotoKase2024}. The more recent ring size measurements discussed in Sec.~\ref{sec:bounds} cannot be inserted as direct measurements of $\theta_\infty$, $s$, or $\mathcal R_{\mathrm{rel}}$~\cite{EHTM872024,EHTM872025}. A smaller uncertainty on the fitted emission ring diameter does not by itself tighten Table~\ref{tab:SBconditional}.

For Sgr~A$^*$, the precession band in Eq.~\eqref{eq:S2band} can also be intersected with the shadow compatible region. At a specified $\zeta$, let $\ell_-(\zeta)$ and $\ell_+(\zeta)$ be the lower and upper endpoints of that band. Thereby, Eq.~\eqref{eq:SDLconicalresponse} implies
\begin{equation}
\frac{1}{\sqrt{1 - \ell_-(\zeta)}}
\lesssim\frac{\Delta\tau_{n + 1,n}(\ell)}{\Delta\tau_{n + 1,n}(0)}
\lesssim\frac{1}{\sqrt{1 - \ell_+(\zeta)}}.
\label{eq:SBprecessiontiming}
\end{equation}
This confines the conical change of the leading delay to a few parts in $10^4$ under the stated orbital assumptions. It is an intersection of conditional summaries. A statistical combination must account for the stellar dynamical information already used in the EHT mass prior~\cite{GRAVITY2024,EHTSgrVI2022}.

\begin{figure}[tbp]
  \centering
  \includegraphics[scale=0.425]{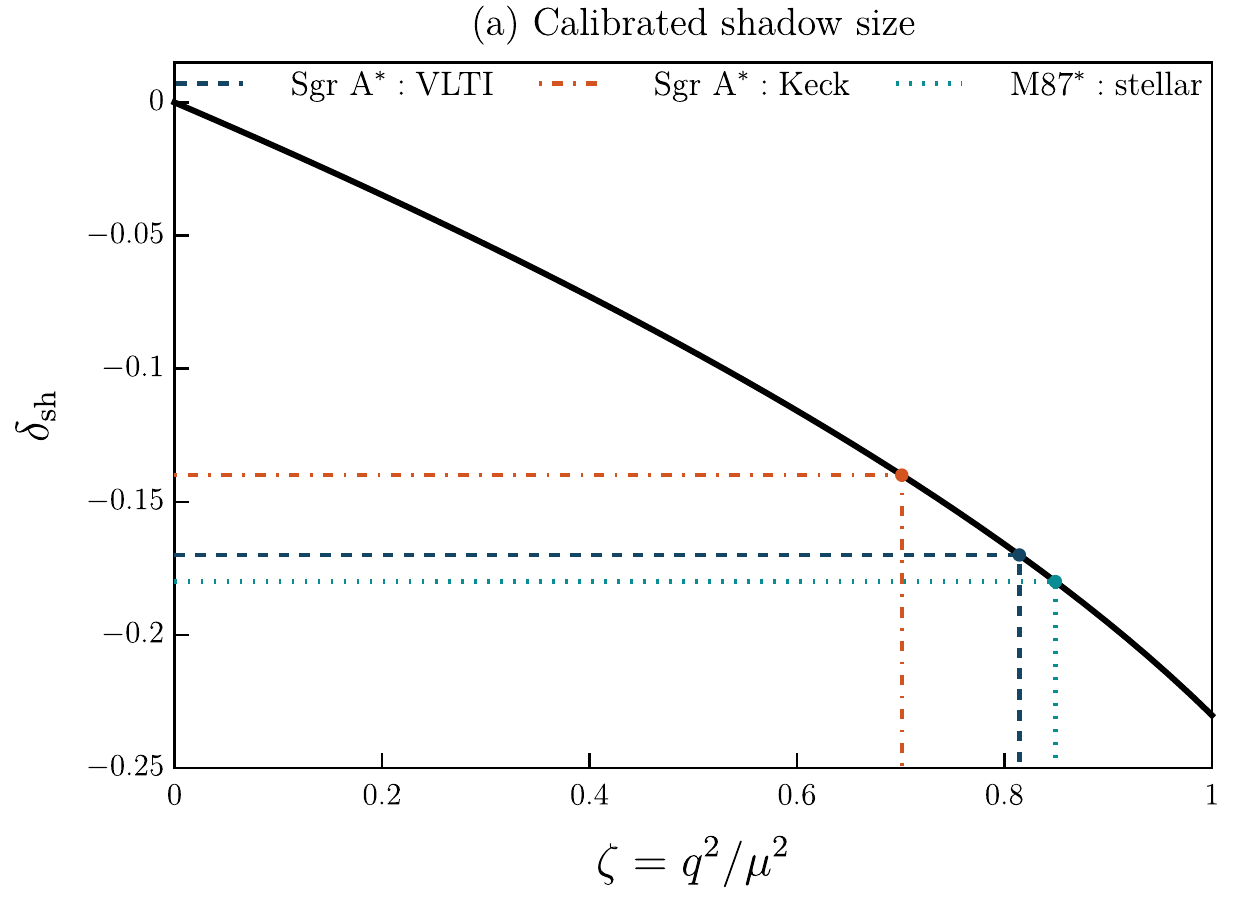}
  \includegraphics[scale=0.425]{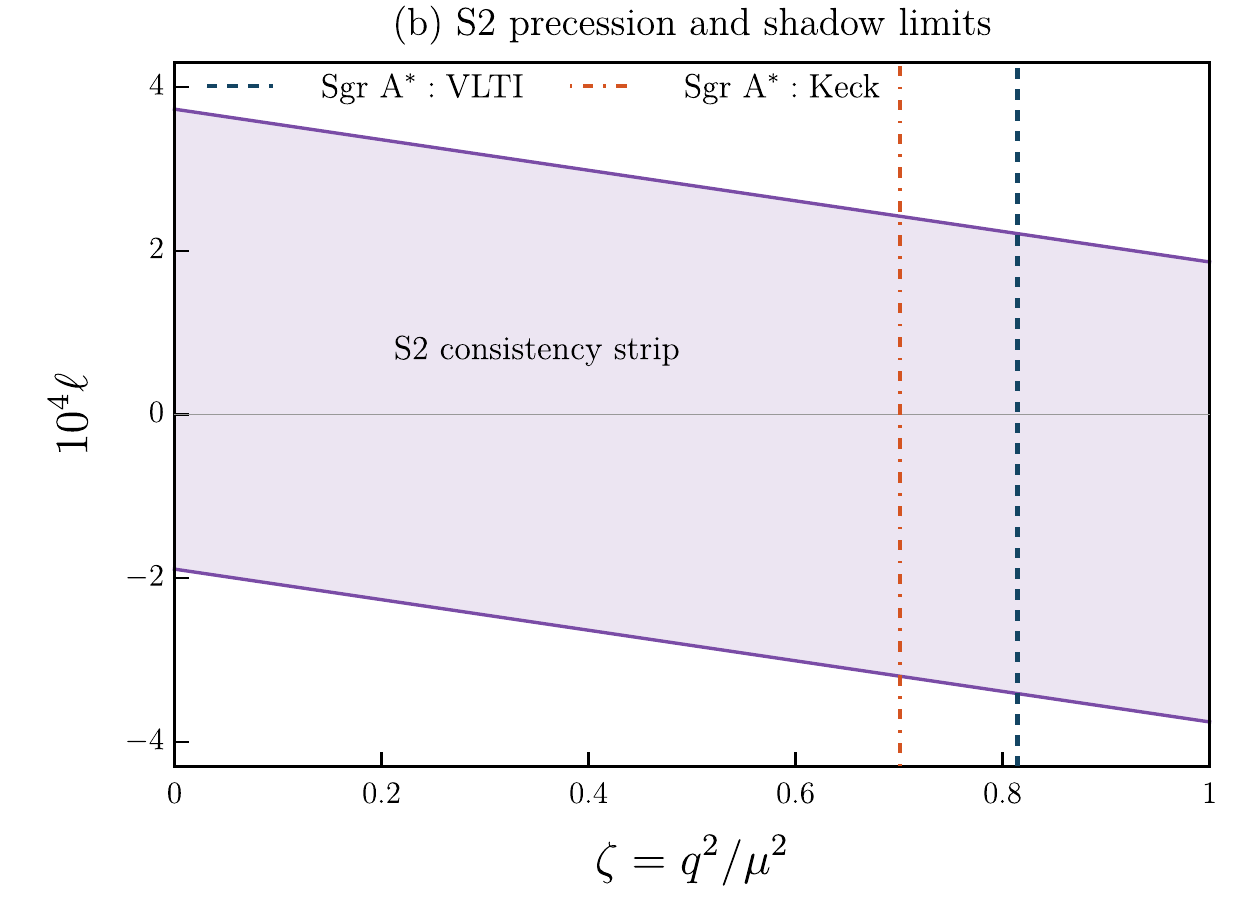}
  \caption{(a) Shadow deviation $\delta_{\mathrm{sh}}(\zeta)$ and the lower endpoints $-0.17$, $-0.14$, and $-0.18$ of the quoted 68\% shadow intervals, giving $\zeta_{\max}=0.8144,0.7011,0.8492$ for Sgr~A$^*$ (VLTI), Sgr~A$^*$ (Keck), and M87$^*$ (stellar dynamics), respectively. (b) The shaded S2 strip satisfies $-1.89\leq10^4\ell + 1.8673\zeta\leq3.73$; the vertical lines show the two alternative Sgr~A$^*$ shadow limits, with smaller charges lying to their left. The M87$^*$ limit is shown  only in (a). }
  \label{fig:kr-bounds}
\end{figure}

Figure~\ref{fig:kr-bounds} illustrates the complementary parameter dependence of shadow size and stellar precession. The shadow limits restrict $\zeta$, whereas the S2 estimate selects a strip in $\ell + \epsilon_\star\zeta$. 

\subsection{Bounds from image separations and relative brightness}

A measured winding hierarchy would constrain the conical parameter more directly. It is convenient to form the dimensionless observable
\begin{equation}
\mathcal G\equiv\frac{\ln(1 + \mathcal R_{\mathrm{rel}})}{2\pi}
\simeq\frac{\sqrt a}{\bar a}.
\label{eq:SBG}
\end{equation}
In an equivalent manner, the decrease of successive image offsets gives $\mathcal G\simeq - \ln\varrho/(2\pi)$ through Eq.~\eqref{eq:SDLhierarchy}.

Suppose a measurement restricts $\mathcal G$ to a positive interval $[\mathcal G_-,\mathcal G_+]$, and an independent or jointly fitted charge restriction gives $0 \leq \zeta \leq \zeta_{\max}$. Since $\bar a^{-2} = \Delta_m$, eliminating $a$ yields the interval enclosure
\begin{equation}
1 - \frac{1}{\mathcal G_-^2}
\lesssim\ell\lesssim
1 - \frac{\Delta_m(\zeta_{\max})}{\mathcal G_+^2},
\qquad \ell<\frac12.
\label{eq:SBfluxellband}
\end{equation}
The full black hole analysis corresponds to $\Delta_m(\zeta_{\max}) = 1/2$. This shows explicitly why a flux ratio alone does not determine $\ell$: the effective charge also changes the demagnification exponent.

The separation gives a second relation that removes this degeneracy within the model. Combining Eqs.~\eqref{eq:SDLobservables} and \eqref{eq:SDLfluxratio}, define
\begin{equation}
\mathcal C_{\mathrm{obs}}
 \equiv \ln\!\left(\frac{s}{\theta_\infty}\right)
 + \frac32\ln( 1 + \mathcal R_{\mathrm{rel}})
\simeq\frac{\bar b + \pi}{\bar a}
\equiv\mathcal C(\zeta).
\label{eq:SBclosure}
\end{equation}
The explicit winding contribution cancels. Then, this combination does not require an external mass or distance calibration at leading order. The relation applies to the nearly aligned image sequence used to define $s$; also, if its source position correction is significant, Eq.~\eqref{eq:SDLpositions} must be fitted before forming the ratio.

The inversion is unique on the black hole. To see this without a charge expansion, let us write
\begin{equation}
w_m = \sqrt{\frac{3\Delta_m}{2(1 + \Delta_m)}}, \qquad
\mathcal C(\zeta) = 
\ln\!\left[
\frac{432}{(1 + w_m)^2}
\left(\frac{\Delta_m}{1 + \Delta_m}\right)^3\right].
\label{eq:SBclosureexplicit}
\end{equation}
Differentiation gives
\begin{equation}
\frac{\mathrm{d}\mathcal C}{\mathrm{d}\Delta_m}
 = \frac{3 + 2w_m}{\Delta_m(1 + \Delta_m)( 1 + w_m)}>0,
\qquad
\frac{\mathrm{d}\Delta_m}{\mathrm{d}\zeta}
=-\frac{2(2 - \Delta_m)^3}{9\Delta_m}<0.
\label{eq:SBclosuremonotonic}
\end{equation}
Therefore a consistency condition for the measured separation and flux contrast is
\begin{equation}
2 \ln \!\left[4(2 - \sqrt2)\right]
\lesssim\mathcal C_{\mathrm{obs}}
\lesssim\ln\!\left[216(7 - 4\sqrt3)\right],
\label{eq:SBclosurerange}
\end{equation}
whose endpoints are approximately $1.70299$ and $2.74136$. A lower observational bound on $\mathcal C_{\mathrm{obs}}$ gives an upper bound on $\zeta$ by solving the single monotonic equation in Eq.~\eqref{eq:SBclosureexplicit}. Once $\zeta$ is obtained, Eq.~\eqref{eq:SDLellreconstruction} determines $\ell$. A significant failure of Eq.~\eqref{eq:SBclosurerange}, after accounting for measurement errors and the neglected lensing terms, would exclude the assumed segment or image model.


\subsection{Timing bounds and an analytic joint reconstruction}

The angular and temporal information can be combined in a different way. Let $\Delta\tau_{\mathrm{circ}}$ denote the leading winding contribution inferred from adjacent images after accounting for the correction in Eq.~\eqref{eq:SDLtimedelay}. With an independent estimate of $R_O$, form
\begin{equation}
\mathcal K \equiv
\frac{c\,\Delta\tau_{\mathrm{circ}}}
{2\pi R_O\sin\theta_\infty} 
\simeq \sqrt a.
\label{eq:SBK}
\end{equation}
The effective charge cancels from this ratio. The cancellation of the local clock factor has already been established in Eq.~\eqref{eq:SDLdistance}.

A positive allowed interval $[\mathcal K_-,\mathcal K_+]$ gives
\begin{equation}
1 - \mathcal K_-^{-2}
\lesssim \ell \lesssim 1 - \mathcal K_+^{-2},
 \qquad 0 < \mathcal K < \sqrt2,
\label{eq:SBtimingellband}
\end{equation}
where the final inequality enforces that $\ell < 1/2$. Thereby, as we could see, a measured winding time and critical angle can constrain $\ell$ without assigning either an electric or a magnetic charge.

Moreover, the ratio of the two observables in Eqs.~\eqref{eq:SBG} and \eqref{eq:SBK} gives the photon orbit instability factor. Let us define
\begin{equation}
\mathcal U \equiv \left(\frac{\mathcal G}{\mathcal K}\right)^2
 \simeq \Delta_m.
\label{eq:SBU}
\end{equation}
The photon sphere relation implies $R_{\mathrm{ph}}/\mu = 3/(2 - \Delta_m)$. Eliminating this radius then gives the analytic reconstruction
\begin{equation}
\zeta \simeq \frac{9 (1 - \mathcal U)}{2 ( 2 -\mathcal U)^2},
\qquad \ell\simeq1 - \mathcal K^{-2},
\qquad \frac12 \leq \mathcal U\leq1.
\label{eq:SBjointinverse}
\end{equation}
In observable space, the charge condition becomes
\begin{equation}
\mathcal G \lesssim \mathcal K\lesssim\sqrt2\,\mathcal G.
\label{eq:SBobservablewedge}
\end{equation}
The uncharged and extremal solutions lie on its two boundaries. This relation is a direct test of the static dyonic geometry; its geometric origin is the connection between the demagnification exponent, the orbit period, and the radial instability~\cite{Stefanov2010,Hadar2021,Salehi2025}.

For interval bounds, let $\mathcal G$ and $\mathcal K$ range over positive intervals as above and set
\begin{equation}
u_- = \max\!\left\{\frac12,
\left(\frac{\mathcal G_-}{\mathcal K_+}\right)^2\right\},\qquad
u_+ = \min\!\left\{1,
\left(\frac{\mathcal G_+}{\mathcal K_-}\right)^2\right\}.
\label{eq:SBuband}
\end{equation}
If $u_-\leq u_+$, monotonicity of Eq.~\eqref{eq:SBjointinverse} gives
\begin{equation}
\frac{9 ( 1 - u_+)}{2 ( 2 - u_+)^2}
\lesssim\zeta\lesssim
\frac{9 ( 1 - u_-)}{ 2 ( 2 - u_-)^2}.
\label{eq:SBchargeband}
\end{equation}
If the intervals do not overlap the physical range, no solution on this part fits both interval restrictions at the adopted approximation order. Eqs.~\eqref{eq:SBuband} and \eqref{eq:SBchargeband} enclose a rectangular data region; correlated measurements should instead be propagated through their joint likelihood. The separate reconstruction from Eq.~\eqref{eq:SBclosure}, in other words, tests whether the measured separation is consistent with the inferred charge.

\begin{figure}[tbp]
  \centering
   \includegraphics[scale=0.435]{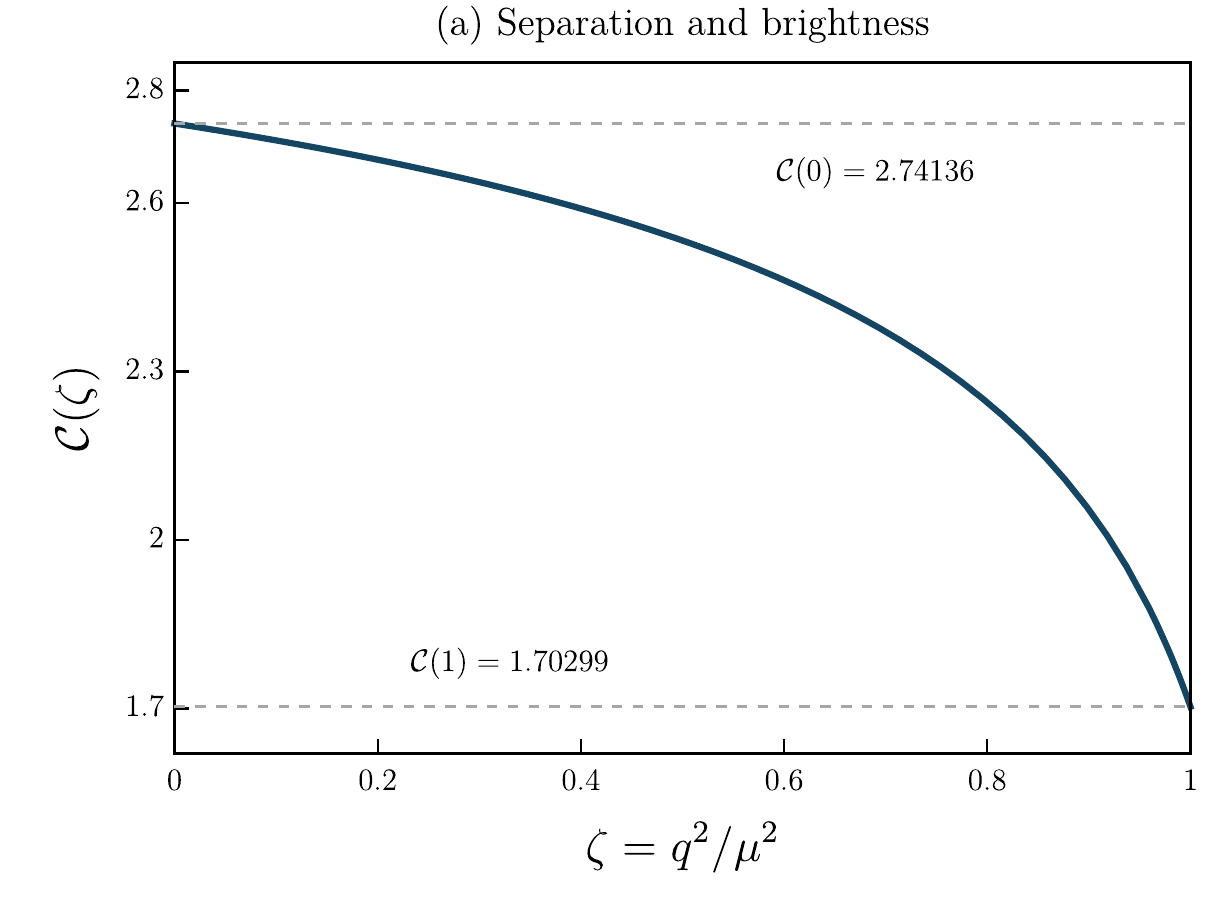}
  \includegraphics[scale=0.438]{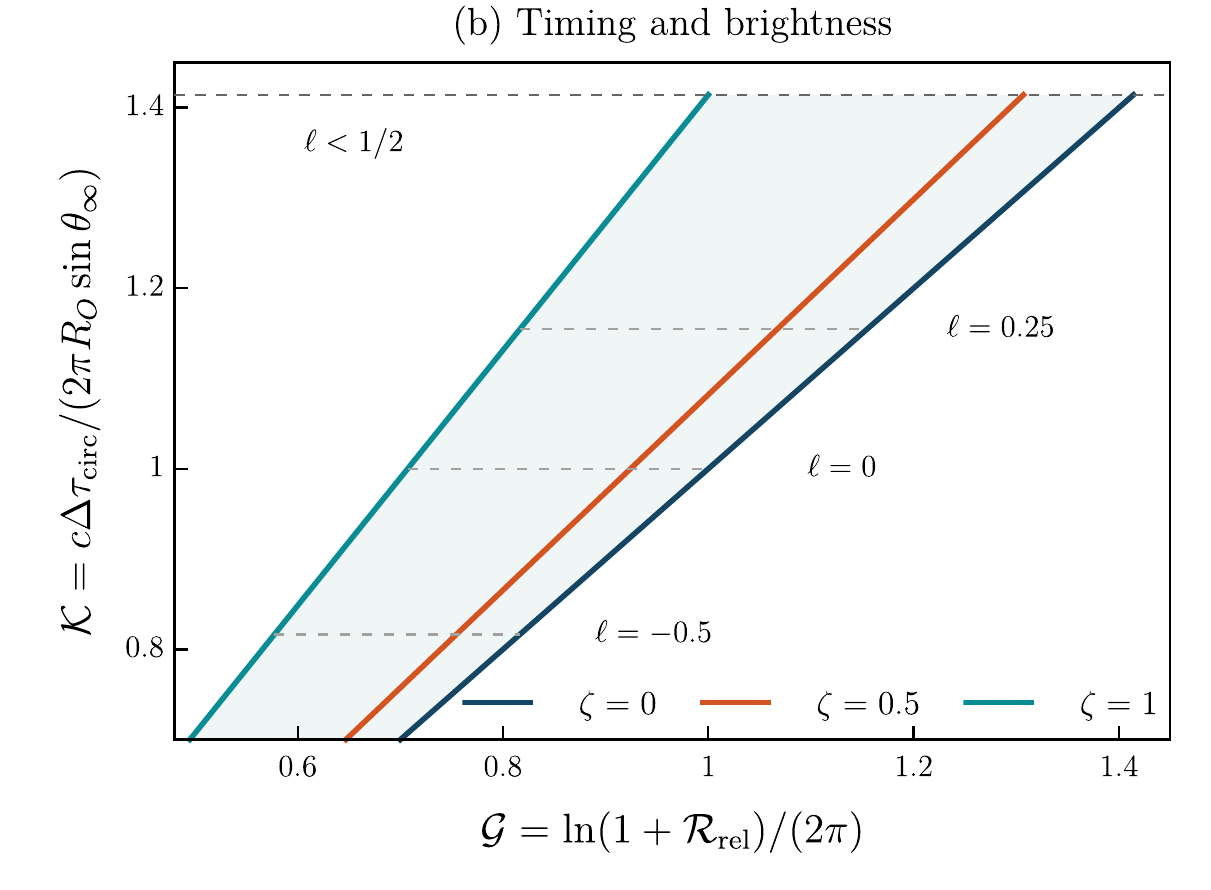}
  \caption{Reconstruction from strong-lensing observables. (a) $\mathcal C=\ln(s/\theta_\infty) + \tfrac32\ln( 1 + \mathcal R_{\mathrm{rel}})$ decreases from $2.74136$ to $1.70299$ over $0\leq\zeta\leq1$. (b) The displayed part of the allowed observable domain satisfies $\mathcal G\leq\mathcal K \leq \sqrt2\mathcal G$ and $0<\mathcal K < \sqrt2$. Coloured lines correspond to $\zeta = 0,0.5,1$; horizontal guides indicate fixed $\ell$. The top dashed boundary, $\ell = 1/2$, is excluded.}
  \label{fig:kr-reconstruction}
\end{figure}

In Fig.~\ref{fig:kr-reconstruction}(a), the monotonic combination $\mathcal C(\zeta)$ gives a unique effective charge within the model. Panel (b) shows the complementary timing reconstruction: $\mathcal K$ fixes $\ell$, while $\mathcal G/\mathcal K$ fixes $\zeta$. The shaded domain represents leading order theoretical consistency; the electric and magnetic charges remain individually degenerate, i.e., in agreement with we have argued before.

\subsection{Sensitivity and requirements for a quantitative bound}

The directions constrained by these observables are transparent near the uncharged, undeformed solution. Expanding at fixed normalized mass and distance gives
\begin{align}
\frac{\theta_\infty}{3\sqrt3\,\theta_g}
& = 1 - \frac{\zeta}{6} + O(\zeta^2),
&\mathcal G & = 1 + \frac{\ell}{2} - \frac{\zeta}{9}
 + O(\ell^2,\ell\zeta,\zeta^2),\nonumber\\
\mathcal K& = 1 + \frac{\ell}{2} + O(\ell^2),
&\mathcal C(\zeta)& = \mathcal C(0)
-\frac{2(3 - \sqrt3)}{9}\zeta + O(\zeta^2).
\label{eq:SBlinearresponse}
\end{align}
The critical radius measures the charge direction, the winding exponent mixes charge and conical deformation, and the timing ratio isolates the latter. In particular,
\begin{equation}
\ell\simeq2(\mathcal K - 1),\qquad
\zeta\simeq9(\mathcal K - \mathcal G).
\label{eq:SBlinearreconstruction}
\end{equation}
For small approximately Gaussian errors about this reference point, the local propagated uncertainties are
\begin{equation}
\sigma_\ell \simeq 2\sigma_{\mathcal K}, \qquad
\sigma_\zeta \simeq
9\sqrt{\sigma_{\mathcal K}^2 + \sigma_{\mathcal G}^2
-2\operatorname{Cov}(\mathcal K,\mathcal G)}.
\label{eq:SBerrors}
\end{equation}
These are sensitivity estimates before applying the boundary $\zeta \geq 0$. For illustration, independent $1\%$ uncertainties on $\mathcal K$ and $\mathcal G$ near unity give $\sigma_\ell\simeq0.020$ and $\sigma_\zeta\simeq0.127$. A sensitivity of order $10^{-4}$ in $\ell$ would require an uncertainty of order $5\times10^{-5}$ in $\mathcal K$.

The distance and angular uncertainties enter that requirement explicitly. Logarithmic differentiation of Eq.~\eqref{eq:SBK} gives
\begin{equation}
\frac{\mathrm{d}\mathcal K}{\mathcal K}
 = \frac{\mathrm{d}\Delta\tau_{\mathrm{circ}}}{\Delta\tau_{\mathrm{circ}}}
 - \frac{\mathrm{d}R_O}{R_O}
 - \cot\theta_\infty\,\mathrm{d}\theta_\infty.
\label{eq:SBtimingerrors}
\end{equation}
At small angles, the last term is $-\mathrm{d}\theta_\infty/\theta_\infty$. For the separation based reconstruction, the analogous error is obtained from
\begin{equation}
\mathrm{d}\mathcal C_{\mathrm{obs}}
 = \frac{\mathrm{d}s}{s}
 - \frac{\mathrm{d}\theta_\infty}{\theta_\infty}
 + \frac{3\,\mathrm{d}\mathcal R_{\mathrm{rel}}}
{2(1 + \mathcal R_{\mathrm{rel}})},
\qquad \sigma_\zeta\simeq
\frac{9\,\sigma_{\mathcal C}}{2(3-\sqrt3)}
\quad(\zeta\simeq0).
\label{eq:SBclosureerrors}
\end{equation}
The covariance of $s$ and $\theta_\infty$ is relevant because $s$ is itself defined relative to the critical curve. Naturally, by omitting it can distort the inferred charge sensitivity.

The accuracy of the lensing approximation must match the proposed measurement. For the first two same side images at $\ell = 0$, near alignment and distant endpoints, Eq.~\eqref{eq:SDLtimedelay} gives the fractional correction to the leading winding delay,
\begin{equation}
\frac{\Delta\tau_{2,1} - \Delta\tau_{\mathrm{circ}}}
{\Delta\tau_{\mathrm{circ}}}
\simeq\frac{\bar a e_1(1 - \varrho)}{2\pi}.
\label{eq:SBtimingbias}
\end{equation}
It is approximately $1.99\times10^{-4}$ for $\zeta = 0$ and $1.56\times10^{-3}$ for $\zeta = 1$. Interpreting the uncorrected delay through Eq.~\eqref{eq:SBtimingellband} would introduce a spurious conical shift at a precision comparable to, or larger than, the precession estimate. The same principle applies to flux and separation bounds: retaining a geometric sum does not remove the higher order bending and source position corrections~\cite{Tsukamoto2023}.

Finally, the observables must be extracted from the emission geometry to which the lens equations apply. A repeated feature can be compared between relativistic images only after accounting for its delay and propagation effects. Correlations of near horizon emission offer a related timing approach, but require an emission model and the appropriate orbit counting convention~\cite{Hadar2021}; accretion flow subrings are commonly indexed by half orbits, whereas $n$ here counts complete windings of a background source ray. Likewise, the total flux of a bright emission ring cannot replace $\mathcal R_{\mathrm{rel}}$. Analyses of existing photon ring claims illustrate this distinction~\cite{LockhartGralla2022,TsukamotoKase2024}.

Long baseline interferometry and time resolved imaging motivate future tests of the image hierarchy~\cite{Johnson2020,LupsascaBHEX2024,Ayzenberg2025}. Their response to spin, inclination, source structure, and propagation must be included when applying the present static predictions to Sgr~A$^*$ or M87$^*$~\cite{Salehi2025,Ayzenberg2025}. A quantitative inference should fit the lensing data and the mass distance calibration together, retaining their covariance and using each prior once. The published shadow intervals provide the conditional bounds in Table~\ref{tab:SBconditional}; Eqs.~\eqref{eq:SBfluxellband}, \eqref{eq:SBclosurerange}, \eqref{eq:SBtimingellband}, and \eqref{eq:SBchargeband} specify the additional constraints that resolved images and differential timing could supply.


\section{Conclusion}
\label{sec:conclusion}

In this work, we investigated gravitational lensing in the weak-- and strong--deflection regimes of a dyonic black hole in Kalb--Ramond gravity. We adopted minimally coupled test radiation and fixed the asymptotic clock, radial scale, and angular convention before the lensing calculation. This normalization separated the locally Reissner--Nordstr\"om form of the equatorial trajectories from the global conical identification. It also exposed the different dependence of the electric and magnetic contributions on the Lorentz--violating parameter: at fixed original charges, the additional magnetic contribution to the normalized squared charge began at order $\ell^2$.

For weak deflection, we evaluated the optical Gaussian curvature and applied the Gauss--Bonnet theorem with the first order displacement of the incoming ray. This correction supplied the contribution required for the complete quadratic mass term. The resulting bending angle contained the second order coefficient $\pi(15\mu^2-3q^2)/(4b^2)$, and independent calculations from the orbit equation, the turning point integral, and Fermat's principle reproduced it. We then obtained the finite distance lens map and derived the image positions, Einstein rings, magnifications, centroid shifts, and differential delays. The construction retained the conical contribution in the source observer geometry and distinguished it from the local curvature induced bending.

For strong deflection, we implemented Tsukamoto's method and evaluated the regular contribution at the photon sphere analytically. Both strong--deflection coefficients retained their exact dependence on the effective charge, and the deflection remained logarithmically divergent throughout the black hole analysis, including extremality. We restored the physical angular identification in the relativistic image sequences and obtained their separations, relative brightness, and arrival time differences. At fixed normalized mass, effective charge, and observer distance, positive $\ell$ moved the first relativistic image closer to the critical curve, increased its flux contrast with the higher order images, and lengthened the winding delay, while the critical angular scale remained unchanged.

The applications to Sgr~A$^*$ and M87$^*$ connected these predictions with the available mass and shadow calibrations. The adopted shadow size intervals yielded upper limits on $\zeta$ of $0.8144$ and $0.7011$ for the alternative Sgr~A$^*$ mass priors and $0.8492$ for the M87$^*$ stellar dynamical approach. These values represented conditional mappings of published intervals, and their propagation restricted the predicted strong lensing observables. The S2 precession estimate supplied a complementary band of order $10^{-4}$ for the conical parameter after the charge contribution had been specified. Throughout these comparisons, we distinguished the critical curve from the bright emission ring and retained the dependence on the adopted mass normalization.

We also obtained two complementary reconstructions from the strong lensing observables. A combination of angular separation and relative brightness cancelled the explicit winding contribution and depended monotonically on the effective charge, without an external mass or distance at leading order. A second combination, formed from the winding delay, the critical angle, and an independently calibrated distance, isolated the conical parameter. Its combination with the demagnification exponent then yielded an analytic reconstruction of both $\ell$ and $\zeta$. The electric and magnetic charges nevertheless remained individually degenerate within the lensing description.

Finally, we quantified the accuracy required for these reconstructions. In the reference configurations, the first exponential correction modified the leading order delay between adjacent images by a fractional amount of $10^{-4}$--$10^{-3}$, and neglecting it mimicked a small conical deformation. Future work could investigate particle creation and scattering effects using the approaches developed in \cite{AraujoFilho:2025rwr,AraujoFilho:2025hkm,AraujoFilho:2024ctw,dePaula:2025kif,Dolan:2024qqr,dePaula:2024xnd}. Another promising direction is to explore the corresponding statistical ensembles through the optical--mechanical analogy \cite{furtado2023thermal,araujo2021bouncing,AraujoFilho:2025fwd,AraujoFilho:2026yaj,araujo2022particles}.


\section*{Acknowledgments}
\hspace{0.5cm} A. A. Araújo Filho is supported by Conselho Nacional de Desenvolvimento Cient\'{\i}fico e Tecnol\'{o}gico (CNPq) -- [150223/2025-0].


\section*{Data Availability Statement}

Data associated with this study consist of the analytical expressions and numerical figures presented in the manuscript. No additional dataset is required to reproduce the analytical results. The code used to generate the results and figures is available from the corresponding author upon reasonable request.

\bibliographystyle{unsrtnat}
\bibliography{main}

\begin{thebibliography}{102}
\providecommand{\natexlab}[1]{#1}
\providecommand{\url}[1]{\texttt{#1}}
\expandafter\ifx\csname urlstyle\endcsname\relax
  \providecommand{\doi}[1]{doi: #1}\else
  \providecommand{\doi}{doi: \begingroup \urlstyle{rm}\Url}\fi

\bibitem[Perlick(2004{\natexlab{a}})]{Perlick2004}
Volker Perlick.
\newblock {On the Exact gravitational lens equation in spherically symmetric
  and static space-times}.
\newblock \emph{Phys. Rev. D}, 69:\penalty0 064017, 2004{\natexlab{a}}.
\newblock \doi{10.1103/PhysRevD.69.064017}.

\bibitem[Perlick(2004{\natexlab{b}})]{PerlickReview2004}
V.~Perlick.
\newblock {Gravitational lensing from a spacetime perspective}.
\newblock \emph{Living Rev. Rel.}, 7:\penalty0 9, 2004{\natexlab{b}}.
\newblock \doi{10.12942/lrr-2004-9}.

\bibitem[Akiyama et~al.(2019)]{EHTM872019}
Kazunori Akiyama et~al.
\newblock {First M87 Event Horizon Telescope Results. VI. The Shadow and Mass
  of the Central Black Hole}.
\newblock \emph{Astrophys. J. Lett.}, 875\penalty0 (1):\penalty0 L6, 2019.
\newblock \doi{10.3847/2041-8213/ab1141}.

\bibitem[Akiyama et~al.(2022{\natexlab{a}})]{EHTSgrI2022}
Kazunori Akiyama et~al.
\newblock {First Sagittarius A* Event Horizon Telescope Results. I. The Shadow
  of the Supermassive Black Hole in the Center of the Milky Way}.
\newblock \emph{Astrophys. J. Lett.}, 930\penalty0 (2):\penalty0 L12,
  2022{\natexlab{a}}.
\newblock \doi{10.3847/2041-8213/ac6674}.

\bibitem[Akiyama et~al.(2022{\natexlab{b}})]{EHTSgrVI2022}
Kazunori Akiyama et~al.
\newblock {First Sagittarius A* Event Horizon Telescope Results. VI. Testing
  the Black Hole Metric}.
\newblock \emph{Astrophys. J. Lett.}, 930\penalty0 (2):\penalty0 L17,
  2022{\natexlab{b}}.
\newblock \doi{10.3847/2041-8213/ac6756}.

\bibitem[Abuter et~al.(2020)]{GRAVITY2020}
R.~Abuter et~al.
\newblock {Detection of the Schwarzschild precession in the orbit of the star
  S2 near the Galactic centre massive black hole}.
\newblock \emph{Astron. Astrophys.}, 636:\penalty0 L5, 2020.
\newblock \doi{10.1051/0004-6361/202037813}.

\bibitem[Abuter et~al.(2022)]{GRAVITY2022}
R.~Abuter et~al.
\newblock {Mass distribution in the Galactic Center based on interferometric
  astrometry of multiple stellar orbits}.
\newblock \emph{Astron. Astrophys.}, 657:\penalty0 L12, 2022.
\newblock \doi{10.1051/0004-6361/202142465}.

\bibitem[Kostelecky and Samuel(1989)]{Kostelecky1989}
V.~Alan Kostelecky and Stuart Samuel.
\newblock {Spontaneous Breaking of Lorentz Symmetry in String Theory}.
\newblock \emph{Phys. Rev. D}, 39:\penalty0 683, 1989.
\newblock \doi{10.1103/PhysRevD.39.683}.

\bibitem[Kostelecky(2004)]{Kostelecky2004}
V.~Alan Kostelecky.
\newblock {Gravity, Lorentz violation, and the standard model}.
\newblock \emph{Phys. Rev. D}, 69:\penalty0 105009, 2004.
\newblock \doi{10.1103/PhysRevD.69.105009}.

\bibitem[Kalb and Ramond(1974)]{Kalb1974}
Michael Kalb and Pierre Ramond.
\newblock {Classical direct interstring action}.
\newblock \emph{Phys. Rev. D}, 9:\penalty0 2273--2284, 1974.
\newblock \doi{10.1103/PhysRevD.9.2273}.

\bibitem[Higashijima and Yokoi(2001)]{Higashijima2001}
Kiyoshi Higashijima and Naoto Yokoi.
\newblock {Spontaneous Lorentz symmetry breaking by antisymmetric tensor
  field}.
\newblock \emph{Phys. Rev. D}, 64:\penalty0 025004, 2001.
\newblock \doi{10.1103/PhysRevD.64.025004}.

\bibitem[Altschul et~al.(2010)Altschul, Bailey, and Kostelecky]{Altschul2010}
Brett Altschul, Quentin~G. Bailey, and V.~Alan Kostelecky.
\newblock {Lorentz violation with an antisymmetric tensor}.
\newblock \emph{Phys. Rev. D}, 81:\penalty0 065028, 2010.
\newblock \doi{10.1103/PhysRevD.81.065028}.

\bibitem[Maluf et~al.(2018)Maluf, Ara{\'u}jo~Filho, Cruz, and
  Almeida]{Maluf2018}
R.~V. Maluf, A.~A. Ara{\'u}jo~Filho, W.~T. Cruz, and C.~A.~S. Almeida.
\newblock {Antisymmetric tensor propagator with spontaneous Lorentz violation}.
\newblock \emph{EPL}, 124\penalty0 (6):\penalty0 61001, 2018.
\newblock \doi{10.1209/0295-5075/124/61001}.

\bibitem[Lessa et~al.(2020)Lessa, Silva, Maluf, and Almeida]{Lessa2020}
L.~A. Lessa, J.~E.~G. Silva, R.~V. Maluf, and C.~A.~S. Almeida.
\newblock {Modified black hole solution with a background
  Kalb{\textendash}Ramond field}.
\newblock \emph{Eur. Phys. J. C}, 80\penalty0 (4):\penalty0 335, 2020.
\newblock \doi{10.1140/epjc/s10052-020-7902-1}.

\bibitem[Yang et~al.(2023)Yang, Chen, Duan, and Zhao]{Yang2023}
Ke~Yang, Yue-Zhe Chen, Zheng-Qiao Duan, and Ju-Ying Zhao.
\newblock {Static and spherically symmetric black holes in gravity with a
  background Kalb-Ramond field}.
\newblock \emph{Phys. Rev. D}, 108\penalty0 (12):\penalty0 124004, 2023.
\newblock \doi{10.1103/PhysRevD.108.124004}.

\bibitem[Duan et~al.(2024)Duan, Zhao, and Yang]{Duan2024}
Zheng-Qiao Duan, Ju-Ying Zhao, and Ke~Yang.
\newblock {Electrically charged black holes in gravity with a background
  Kalb{\textendash}Ramond field}.
\newblock \emph{Eur. Phys. J. C}, 84\penalty0 (8):\penalty0 798, 2024.
\newblock \doi{10.1140/epjc/s10052-024-13188-5}.

\bibitem[Liu et~al.(2024)Liu, Wu, and Wang]{Liu2024}
Wentao Liu, Di~Wu, and Jieci Wang.
\newblock {Static neutral black holes in Kalb-Ramond gravity}.
\newblock \emph{JCAP}, 09:\penalty0 017, 2024.
\newblock \doi{10.1088/1475-7516/2024/09/017}.

\bibitem[Liu et~al.(2025)Liu, Wu, Wei, and Liu]{Liu2025}
Jia-Zhou Liu, Shan-Ping Wu, Shao-Wen Wei, and Yu-Xiao Liu.
\newblock {Exact black hole solutions in gravity with a background Kalb-Ramond
  field}.
\newblock \emph{JCAP}, 11:\penalty0 056, 2025.
\newblock \doi{10.1088/1475-7516/2025/11/056}.

\bibitem[Ara{\'u}jo~Filho et~al.(2025{\natexlab{a}})Ara{\'u}jo~Filho, Heidari,
  and Lobo]{AraujoFilho:2025jcu}
A.~A. Ara{\'u}jo~Filho, N.~Heidari, and Iarley~P. Lobo.
\newblock {A non-commutative Kalb-Ramond black hole}.
\newblock \emph{JCAP}, 09:\penalty0 076, 2025{\natexlab{a}}.
\newblock \doi{10.1088/1475-7516/2025/09/076}.

\bibitem[Yang et~al.(2026)Yang, Guo, Liu, and Liu]{Yang2026}
Jia-Hui Yang, Xin-Yu Guo, Jia-Zhou Liu, and Yu-Xiao Liu.
\newblock {Charged Black Holes with a Lorentz--Violating Kalb--Ramond
  Background}.
\newblock arXiv:2608.02196 [gr-qc], 8 2026.
\newblock URL \url{https://arxiv.org/abs/2608.02196}.
\newblock Preprint.

\bibitem[Lin et~al.(2026)Lin, Liu, and Liu]{Lin2026}
Yu-Xuan Lin, Jia-Zhou Liu, and Yu-Xiao Liu.
\newblock {Dyonic Black Holes in Lorentz-Violating Gravity with a Background
  Kalb--Ramond Field}.
\newblock arXiv:2605.18371 [gr-qc], 5 2026.
\newblock URL \url{https://arxiv.org/abs/2605.18371}.
\newblock Preprint.

\bibitem[Ara{\'u}jo~Filho et~al.(2024)Ara{\'u}jo~Filho, Reis, and
  Hassanabadi]{Ara2024}
A.~A. Ara{\'u}jo~Filho, J.~A. A.~S. Reis, and H.~Hassanabadi.
\newblock {Exploring antisymmetric tensor effects on black hole shadows and
  quasinormal frequencies}.
\newblock \emph{JCAP}, 05:\penalty0 029, 2024.
\newblock \doi{10.1088/1475-7516/2024/05/029}.

\bibitem[Ara{\'u}jo~Filho et~al.(2025{\natexlab{b}})Ara{\'u}jo~Filho, Heidari,
  Reis, and Hassanabadi]{AraCharged2025}
A.~A. Ara{\'u}jo~Filho, N.~Heidari, J.~A. A.~S. Reis, and H.~Hassanabadi.
\newblock {The impact of an antisymmetric tensor on charged black holes:
  evaporation process, geodesics, deflection angle, scattering effects and
  quasinormal modes}.
\newblock \emph{Class. Quant. Grav.}, 42\penalty0 (6):\penalty0 065026,
  2025{\natexlab{b}}.
\newblock \doi{10.1088/1361-6382/adbb4f}.

\bibitem[Shi et~al.(2026)Shi, Ara{\'u}jo~Filho, de~Farias, Bezerra, and
  Queiroz]{Shi:2025xkd}
Yuxuan Shi, A.~A. Ara{\'u}jo~Filho, K.~E.~L. de~Farias, V.~B. Bezerra, and
  Amilcar~R. Queiroz.
\newblock {Neutrino oscillations in a Kalb-Ramond black hole background}.
\newblock \emph{Eur. Phys. J. Plus}, 141\penalty0 (9):\penalty0 1038, 2026.
\newblock \doi{10.1140/epjp/s13360-026-08255-7}.

\bibitem[Shi and Ara{\'u}jo~Filho(2025)]{Shi:2025rfq}
Yuxuan Shi and A.~A. Ara{\'u}jo~Filho.
\newblock {Influence of a Kalb-Ramond black hole on neutrino behavior}.
\newblock \emph{JHEP}, 08:\penalty0 028, 2025.
\newblock \doi{10.1007/JHEP08(2025)028}.

\bibitem[Rodrigues et~al.(2026)Rodrigues, Lobo, and
  Rodrigues]{Rodrigues:2026ofx}
Luiz F.~G. Rodrigues, Francisco S.~N. Lobo, and Manuel~E. Rodrigues.
\newblock {Lorentz-violating signatures in quasi-periodic oscillations from a
  magnetised Kalb-Ramond black hole}.
\newblock \emph{JCAP}, 09:\penalty0 062, 2026.
\newblock \doi{10.1088/1475-7516/2026/09/062}.

\bibitem[Belchior et~al.(2025{\natexlab{a}})Belchior, Maluf, Petrov, and
  Porf{\'\i}rio]{BelchiorMonopole2025}
Fernando~M. Belchior, Roberto~V. Maluf, Albert~Yu. Petrov, and Paulo~J.
  Porf{\'\i}rio.
\newblock {Global monopole in a Ricci-coupled Kalb--Ramond bumblebee gravity}.
\newblock \emph{Eur. Phys. J. C}, 85\penalty0 (6):\penalty0 658,
  2025{\natexlab{a}}.
\newblock \doi{10.1140/epjc/s10052-025-14390-9}.

\bibitem[Al-Badawi et~al.(2025)Al-Badawi, Ahmed, and
  Sakall{\i}]{AhmedModMax2025}
Ahmad Al-Badawi, Faizuddin Ahmed, and {\.I}zzet Sakall{\i}.
\newblock {Particle dynamics and thermal properties in Kalb--Ramond ModMax
  black holes: Theoretical predictions for observational tests of exotic
  physics}.
\newblock \emph{Phys. Dark Univ.}, 50:\penalty0 102076, 2025.
\newblock \doi{10.1016/j.dark.2025.102076}.

\bibitem[Sekhmani et~al.(2025{\natexlab{a}})Sekhmani, Baruah, Maurya,
  Rayimbaev, Altanji, Ibragimov, and Muminov]{SekhmaniPhantom2025}
Y.~Sekhmani, A.~Baruah, S.~K. Maurya, J.~Rayimbaev, M.~Altanji, I.~Ibragimov,
  and S.~Muminov.
\newblock {Kalb-Ramond black holes sourced by ModMax electrodynamics: Some
  perturbative properties in the phantom sector}.
\newblock \emph{Phys. Dark Univ.}, 50:\penalty0 102157, 2025{\natexlab{a}}.
\newblock \doi{10.1016/j.dark.2025.102157}.

\bibitem[Sucu et~al.(2026)Sucu, Sakall{\i}, and Saridakis]{SucuBranches2026}
Erdem Sucu, {\.I}zzet Sakall{\i}, and Emmanuel~N. Saridakis.
\newblock {Branch structure and nonextensive thermodynamics of
  Kalb-Ramond-ModMax black holes: Observational signatures}.
\newblock \emph{Phys. Rev. D}, 114\penalty0 (4):\penalty0 044006, 2026.
\newblock \doi{10.1103/mq5t-7sj4}.

\bibitem[Belchior et~al.(2026)Belchior, Ahmed, and Silva]{BelchiorModMax2026}
Fernando~M. Belchior, Faizuddin Ahmed, and Edilberto~O. Silva.
\newblock {ModMax black hole surrounded by perfect-fluid dark matter in
  Lorentz-violating Kalb-Ramond gravity}.
\newblock arXiv:2605.26131 [gr-qc], May 2026.
\newblock URL \url{https://arxiv.org/abs/2605.26131}.
\newblock Preprint.

\bibitem[Murodov(2026)]{MurodovWormhole2026}
Sardor Murodov.
\newblock {Slow rotation of a localized Kalb-Ramond wormhole: Wald charges and
  fractional quadrupolar hair}.
\newblock arXiv:2609.05921 [gr-qc], September 2026.
\newblock URL \url{https://arxiv.org/abs/2609.05921}.
\newblock Preprint.

\bibitem[Murodov et~al.(2026{\natexlab{a}})Murodov, Kholturayev, Ahmedov,
  Rahmatov, Egamberdiev, and Ahmedov]{MurodovDirac2026}
Sardor Murodov, Olimjon Kholturayev, Bahodir Ahmedov, Bekzod Rahmatov, Islom
  Egamberdiev, and Bobomurat Ahmedov.
\newblock {Massive neutral Dirac quasibound states in a Newman-Janis-generated
  rotating charged Kalb-Ramond black-hole geometry}.
\newblock arXiv:2608.09313 [gr-qc], August 2026{\natexlab{a}}.
\newblock URL \url{https://arxiv.org/abs/2608.09313}.
\newblock Preprint.

\bibitem[Murodov et~al.(2026{\natexlab{b}})Murodov, Kholturayev, Rahmatov,
  Rayimbaev, Egamberdiev, and Karshiboev]{MurodovBZ2026}
Sardor Murodov, Olimjon Kholturayev, Bekzod Rahmatov, Javlon Rayimbaev, Islom
  Egamberdiev, and Shavkat Karshiboev.
\newblock {Blandford-Znajek Scaling in a Power-Law Rotating Kalb-Ramond
  Geometry: Magnetic-Flux Systematics and Bayesian Identifiability}.
\newblock arXiv:2608.11962 [gr-qc], August 2026{\natexlab{b}}.
\newblock URL \url{https://arxiv.org/abs/2608.11962}.
\newblock Preprint.

\bibitem[Rahmatov et~al.(2026)Rahmatov, Rahmonova, Murodov, Rayimbaev, and
  Ahmedov]{Rahmatov2026}
Bekzod Rahmatov, Shahnoza Rahmonova, Sardor Murodov, Javlon Rayimbaev, and
  Bobomurat Ahmedov.
\newblock {Strong gravitational lensing by charged Kalb--Ramond black holes
  with two nonminimal couplings}.
\newblock \emph{Eur. Phys. J. Plus}, 141:\penalty0 1051, 2026.
\newblock \doi{10.1140/epjp/s13360-026-08279-z}.
\newblock URL
  \url{https://link.springer.com/article/10.1140/epjp/s13360-026-08279-z}.

\bibitem[Gibbons and Werner(2008)]{GibbonsWerner2008}
G.~W. Gibbons and M.~C. Werner.
\newblock {Applications of the Gauss-Bonnet theorem to gravitational lensing}.
\newblock \emph{Class. Quant. Grav.}, 25:\penalty0 235009, 2008.
\newblock \doi{10.1088/0264-9381/25/23/235009}.

\bibitem[Werner(2012)]{Werner2012}
M.~C. Werner.
\newblock {Gravitational lensing in the Kerr-Randers optical geometry}.
\newblock \emph{Gen. Rel. Grav.}, 44:\penalty0 3047--3057, 2012.
\newblock \doi{10.1007/s10714-012-1458-9}.

\bibitem[Ishihara et~al.(2016)Ishihara, Suzuki, Ono, Kitamura, and
  Asada]{Ishihara2016}
Asahi Ishihara, Yusuke Suzuki, Toshiaki Ono, Takao Kitamura, and Hideki Asada.
\newblock {Gravitational bending angle of light for finite distance and the
  Gauss-Bonnet theorem}.
\newblock \emph{Phys. Rev. D}, 94\penalty0 (8):\penalty0 084015, 2016.
\newblock \doi{10.1103/PhysRevD.94.084015}.

\bibitem[Ishihara et~al.(2017)Ishihara, Suzuki, Ono, and Asada]{Ishihara2017}
Asahi Ishihara, Yusuke Suzuki, Toshiaki Ono, and Hideki Asada.
\newblock {Finite-distance corrections to the gravitational bending angle of
  light in the strong deflection limit}.
\newblock \emph{Phys. Rev. D}, 95\penalty0 (4):\penalty0 044017, 2017.
\newblock \doi{10.1103/PhysRevD.95.044017}.

\bibitem[Ono and Asada(2019)]{OnoAsada2019}
Toshiaki Ono and Hideki Asada.
\newblock {The effects of finite distance on the gravitational deflection angle
  of light}.
\newblock \emph{Universe}, 5\penalty0 (11):\penalty0 218, 2019.
\newblock \doi{10.3390/universe5110218}.

\bibitem[Keeton and Petters(2005)]{KeetonPetters2005}
Charles~R. Keeton and A.~O. Petters.
\newblock {Formalism for testing theories of gravity using lensing by compact
  objects. I. Static, spherically symmetric case}.
\newblock \emph{Phys. Rev. D}, 72:\penalty0 104006, 2005.
\newblock \doi{10.1103/PhysRevD.72.104006}.

\bibitem[Keeton and Petters(2006)]{KeetonPetters2006}
Charles~R. Keeton and A.~O. Petters.
\newblock {Formalism for testing theories of gravity using lensing by compact
  objects. II. Probing post-post-Newtonian metrics}.
\newblock \emph{Phys. Rev. D}, 73:\penalty0 044024, 2006.
\newblock \doi{10.1103/PhysRevD.73.044024}.

\bibitem[Ara{\'u}jo~Filho et~al.(2025{\natexlab{c}})Ara{\'u}jo~Filho, Heidari,
  and {\"O}vg{\"u}n]{AraNoncomm2025}
A.~A. Ara{\'u}jo~Filho, N.~Heidari, and Ali {\"O}vg{\"u}n.
\newblock {Geodesics, accretion disk, gravitational lensing, time delay, and
  effects on neutrinos induced by a non-commutative black hole}.
\newblock \emph{JCAP}, 06:\penalty0 062, 2025{\natexlab{c}}.
\newblock \doi{10.1088/1475-7516/2025/06/062}.

\bibitem[Jha(2025)]{Jha2025}
Sohan~Kumar Jha.
\newblock {Shadow, ISCO, quasinormal modes, Hawking spectrum, weak
  gravitational lensing, and parameter estimation of a Schwarzschild black hole
  surrounded by a Dehnen type dark matter halo}.
\newblock \emph{JCAP}, 03:\penalty0 054, 2025.
\newblock \doi{10.1088/1475-7516/2025/03/054}.

\bibitem[Sereno(2004)]{Sereno2004}
Mauro Sereno.
\newblock {Weak field limit of Reissner-Nordstrom black hole lensing}.
\newblock \emph{Phys. Rev. D}, 69:\penalty0 023002, 2004.
\newblock \doi{10.1103/PhysRevD.69.023002}.

\bibitem[Iyer and Petters(2007)]{IyerPetters2007}
Savitri~V. Iyer and Arlie~O. Petters.
\newblock {Light's bending angle due to black holes: From the photon sphere to
  infinity}.
\newblock \emph{Gen. Rel. Grav.}, 39:\penalty0 1563--1582, 2007.
\newblock \doi{10.1007/s10714-007-0481-8}.

\bibitem[Atamurotov et~al.(2022)Atamurotov, Ortiqboev, Abdujabbarov, and
  Mustafa]{Atamurotov2022}
Farruh Atamurotov, Dilmurod Ortiqboev, Ahmadjon Abdujabbarov, and G.~Mustafa.
\newblock {Particle dynamics and gravitational weak lensing around black hole
  in the Kalb-Ramond gravity}.
\newblock \emph{Eur. Phys. J. C}, 82\penalty0 (8):\penalty0 659, 2022.
\newblock \doi{10.1140/epjc/s10052-022-10619-z}.

\bibitem[Pantig et~al.(2025)Pantig, {\"O}vg{\"u}n, and Rinc{\'o}n]{Pantig2025}
Reggie~C. Pantig, Ali {\"O}vg{\"u}n, and {\'A}ngel Rinc{\'o}n.
\newblock {Charged black holes in KR gravity: Weak deflection angle, shadow
  cast, quasinormal modes and neutrino annihilation}.
\newblock \emph{Phys. Dark Univ.}, 49:\penalty0 102029, 2025.
\newblock \doi{10.1016/j.dark.2025.102029}.

\bibitem[Pantig and {\"O}vg{\"u}n(2025)]{PantigOvgun2025}
Reggie~C. Pantig and Ali {\"O}vg{\"u}n.
\newblock {Multimodal signatures of asymptotic (A)dS Kalb--Ramond black holes:
  Constraints through the shadow, weak deflection angle, and topological photon
  spheres}.
\newblock \emph{Annals Phys.}, 480:\penalty0 170104, 2025.
\newblock \doi{10.1016/j.aop.2025.170104}.

\bibitem[Filho(2025)]{Filho:2024tgy}
A.~A.~Ara{\'u}jo Filho.
\newblock {Antisymmetric tensor influence on charged black hole lensing
  phenomena and time delay}.
\newblock \emph{JHEAp}, 47:\penalty0 100401, 2025.
\newblock \doi{10.1016/j.jheap.2025.100401}.

\bibitem[Ara{\'u}jo~Filho et~al.(2026{\natexlab{a}})Ara{\'u}jo~Filho, Heidari,
  Lobo, and Shi]{AraujoFilho:2025huk}
A.~A. Ara{\'u}jo~Filho, N.~Heidari, Iarley~P. Lobo, and Yuxuan Shi.
\newblock {Optical phenomena in a non-commutative Kalb{\textendash}Ramond black
  hole spacetime}.
\newblock \emph{Annals Phys.}, 492:\penalty0 170550, 2026{\natexlab{a}}.
\newblock \doi{10.1016/j.aop.2026.170550}.

\bibitem[Sucu and Sakall{\i}(2025)]{SucuSakalli2025}
Erdem Sucu and {\.I}zzet Sakall{\i}.
\newblock {Exploring Lorentz-violating effects of Kalb-Ramond field on charged
  black hole thermodynamics and photon dynamics}.
\newblock \emph{Phys. Rev. D}, 111\penalty0 (6):\penalty0 064049, 2025.
\newblock \doi{10.1103/PhysRevD.111.064049}.

\bibitem[Ahmed et~al.(2025)Ahmed, Sakall{\i}, and Al-Badawi]{AhmedStrings2025}
Faizuddin Ahmed, {\.I}zzet Sakall{\i}, and Ahmad Al-Badawi.
\newblock {Photon Deflection and Magnification in Kalb-Ramond Black Holes with
  Topological String Configurations}.
\newblock arXiv:2507.22673 [gr-qc], July 2025.
\newblock URL \url{https://arxiv.org/abs/2507.22673}.
\newblock Preprint.

\bibitem[{\"O}vg{\"u}n(2025)]{Ovgun2025}
Ali {\"O}vg{\"u}n.
\newblock {Weak gravitational lensing in Ricci-coupled Kalb{\textendash}Ramond
  bumblebee gravity: Global monopole and axion-plasmon medium effects}.
\newblock \emph{Phys. Dark Univ.}, 48:\penalty0 101905, 2025.
\newblock \doi{10.1016/j.dark.2025.101905}.

\bibitem[{\"O}vg{\"u}n et~al.(2026){\"O}vg{\"u}n, Pantig, and
  Panotopoulos]{Ovgun2026}
Ali {\"O}vg{\"u}n, Reggie~C. Pantig, and Grigoris Panotopoulos.
\newblock {Chromatic Weak Lensing by Charged Black Holes with Two
  Lorentz-Violating Kalb-Ramond Couplings}.
\newblock arXiv:2608.18315 [gr-qc], 8 2026.
\newblock URL \url{https://arxiv.org/abs/2608.18315}.
\newblock Preprint.

\bibitem[Virbhadra and Ellis(2000)]{VirbhadraEllis2000}
K.~S. Virbhadra and George F.~R. Ellis.
\newblock {Schwarzschild black hole lensing}.
\newblock \emph{Phys. Rev. D}, 62:\penalty0 084003, 2000.
\newblock \doi{10.1103/PhysRevD.62.084003}.

\bibitem[Bozza et~al.(2001)Bozza, Capozziello, Iovane, and
  Scarpetta]{Bozza2001}
V.~Bozza, S.~Capozziello, G.~Iovane, and G.~Scarpetta.
\newblock {Strong field limit of black hole gravitational lensing}.
\newblock \emph{Gen. Rel. Grav.}, 33:\penalty0 1535--1548, 2001.
\newblock \doi{10.1023/A:1012292927358}.

\bibitem[Bozza(2002)]{Bozza2002}
V.~Bozza.
\newblock {Gravitational lensing in the strong field limit}.
\newblock \emph{Phys. Rev. D}, 66:\penalty0 103001, 2002.
\newblock \doi{10.1103/PhysRevD.66.103001}.

\bibitem[Tsukamoto(2017)]{Tsukamoto2017}
Naoki Tsukamoto.
\newblock {Deflection angle in the strong deflection limit in a general
  asymptotically flat, static, spherically symmetric spacetime}.
\newblock \emph{Phys. Rev. D}, 95\penalty0 (6):\penalty0 064035, 2017.
\newblock \doi{10.1103/PhysRevD.95.064035}.

\bibitem[Eiroa et~al.(2002)Eiroa, Romero, and Torres]{Eiroa2002}
Ernesto~F. Eiroa, Gustavo~E. Romero, and Diego~F. Torres.
\newblock {Reissner-Nordstrom black hole lensing}.
\newblock \emph{Phys. Rev. D}, 66:\penalty0 024010, 2002.
\newblock \doi{10.1103/PhysRevD.66.024010}.

\bibitem[Tsukamoto and Gong(2017)]{TsukamotoGong2017}
Naoki Tsukamoto and Yungui Gong.
\newblock {Retrolensing by a charged black hole}.
\newblock \emph{Phys. Rev. D}, 95\penalty0 (6):\penalty0 064034, 2017.
\newblock \doi{10.1103/PhysRevD.95.064034}.

\bibitem[Bozza and Scarpetta(2007)]{BozzaScarpetta2007}
V.~Bozza and G.~Scarpetta.
\newblock {Strong deflection limit of black hole gravitational lensing with
  arbitrary source distances}.
\newblock \emph{Phys. Rev. D}, 76:\penalty0 083008, 2007.
\newblock \doi{10.1103/PhysRevD.76.083008}.

\bibitem[Bozza and Mancini(2004)]{BozzaMancini2004}
V.~Bozza and L.~Mancini.
\newblock {Time delay in black hole gravitational lensing as a distance
  estimator}.
\newblock \emph{Gen. Rel. Grav.}, 36:\penalty0 435--450, 2004.
\newblock \doi{10.1023/B:GERG.0000010486.58026.4f}.

\bibitem[Belchior et~al.(2025{\natexlab{b}})Belchior, Maluf, Oliveira, Petrov,
  and Porf{\'\i}rio]{BelchiorLensing2025}
F.~M. Belchior, R.~V. Maluf, A.~R.~M. Oliveira, A.~Yu. Petrov, and P.~J.
  Porf{\'\i}rio.
\newblock {Fermionic greybody factors and strong gravitational lensing by
  Lorentz-violating global monopole}.
\newblock arXiv:2508.14861 [gr-qc], August 2025{\natexlab{b}}.
\newblock URL \url{https://arxiv.org/abs/2508.14861}.
\newblock Preprint.

\bibitem[Pereira et~al.(2026)Pereira, Silva, Soares, Ara{\'u}jo~Filho,
  Vit{\'o}ria, and Belich]{PereiraKR2026}
C.~F.~S. Pereira, Marcos V. de~S. Silva, A.~R. Soares, A.~A. Ara{\'u}jo~Filho,
  R.~L.~L. Vit{\'o}ria, and H.~Belich.
\newblock {Light propagation and gravitational lensing effects in charged
  Kalb{\textendash}Ramond spacetime in nonlinear electrodynamics}.
\newblock \emph{Phys. Lett. B}, 879:\penalty0 140699, 2026.
\newblock \doi{10.1016/j.physletb.2026.140699}.

\bibitem[Sekhmani et~al.(2026)Sekhmani, Al-Badawi, Fathi, Vachher, and
  Ghosh]{SekhmaniAnisotropic2026}
Y.~Sekhmani, A.~Al-Badawi, Mohsen Fathi, A.~Vachher, and Sushant~G. Ghosh.
\newblock {Black hole solutions surrounded by an anisotropic fluid in a
  Kalb--Ramond two--form background}.
\newblock arXiv:2603.07052 [gr-qc], March 2026.
\newblock URL \url{https://arxiv.org/abs/2603.07052}.
\newblock Preprint.

\bibitem[Gralla et~al.(2019)Gralla, Holz, and Wald]{Gralla2019}
Samuel~E. Gralla, Daniel~E. Holz, and Robert~M. Wald.
\newblock {Black Hole Shadows, Photon Rings, and Lensing Rings}.
\newblock \emph{Phys. Rev. D}, 100\penalty0 (2):\penalty0 024018, 2019.
\newblock \doi{10.1103/PhysRevD.100.024018}.

\bibitem[Perlick and Tsupko(2022)]{PerlickTsupko2022}
Volker Perlick and Oleg~Yu. Tsupko.
\newblock {Calculating black hole shadows: Review of analytical studies}.
\newblock \emph{Phys. Rept.}, 947:\penalty0 1--39, 2022.
\newblock \doi{10.1016/j.physrep.2021.10.004}.

\bibitem[Tsukamoto and Kase(2024)]{TsukamotoKase2024}
Naoki Tsukamoto and Ryotaro Kase.
\newblock {Constraints on the black-hole charges of M87* and Sagittarius A* by
  changing rates of photon spheres can be relaxed}.
\newblock \emph{Phys. Rev. D}, 110\penalty0 (4):\penalty0 044065, 2024.
\newblock \doi{10.1103/PhysRevD.110.044065}.

\bibitem[Sekhmani et~al.(2025{\natexlab{b}})Sekhmani, Boshkayev,
  Azreg-A{\"\i}nou, Maurya, Altanji, and Urazalina]{SekhmaniRotating2025}
Yassine Sekhmani, Kuantay Boshkayev, Mustapha Azreg-A{\"\i}nou, Sunil~K.
  Maurya, Mohamed Altanji, and Ainur Urazalina.
\newblock {Constraints on Kalb-Ramond Gravity from EHT Observations of Rotating
  Black Holes in Traceless Conformal Electrodynamics}.
\newblock arXiv:2509.16782 [gr-qc], September 2025{\natexlab{b}}.
\newblock URL \url{https://arxiv.org/abs/2509.16782}.
\newblock Preprint.

\bibitem[Johnson et~al.(2020)]{Johnson2020}
Michael~D. Johnson et~al.
\newblock {Universal interferometric signatures of a black
  hole{\textquoteright}s photon ring}.
\newblock \emph{Sci. Adv.}, 6\penalty0 (12):\penalty0 eaaz1310, 2020.
\newblock \doi{10.1126/sciadv.aaz1310}.

\bibitem[Hadar et~al.(2021)Hadar, Johnson, Lupsasca, and Wong]{Hadar2021}
Shahar Hadar, Michael~D. Johnson, Alexandru Lupsasca, and George~N. Wong.
\newblock {Photon Ring Autocorrelations}.
\newblock \emph{Phys. Rev. D}, 103\penalty0 (10):\penalty0 104038, 2021.
\newblock \doi{10.1103/PhysRevD.103.104038}.

\bibitem[Ayzenberg et~al.(2025)]{Ayzenberg2025}
D.~Ayzenberg et~al.
\newblock {Fundamental physics opportunities with future ground-based mm/sub-mm
  VLBI arrays}.
\newblock \emph{Living Rev. Rel.}, 28\penalty0 (1):\penalty0 4, 2025.
\newblock \doi{10.1007/s41114-025-00057-0}.
\newblock [Erratum: Living Rev.Rel. 28, 7 (2025)].

\bibitem[Lupsasca et~al.(2024)Lupsasca, C{\'a}rdenas-Avenda{\~n}o, Palumbo,
  Johnson, Gralla, Marrone, Galison, Tiede, and Keeble]{LupsascaBHEX2024}
Alexandru Lupsasca, Alejandro C{\'a}rdenas-Avenda{\~n}o, Daniel C.~M. Palumbo,
  Michael~D. Johnson, Samuel~E. Gralla, Daniel~P. Marrone, Peter Galison, Paul
  Tiede, and Lennox Keeble.
\newblock {The Black Hole Explorer: photon ring science, detection, and shape
  measurement}.
\newblock \emph{Proc. SPIE Int. Soc. Opt. Eng.}, 13092:\penalty0 130926Q, 2024.
\newblock \doi{10.1117/12.3019437}.

\bibitem[Ara{\'u}jo~Filho(2026{\natexlab{a}})]{AraujoFilho:2026tsc}
A.~A. Ara{\'u}jo~Filho.
\newblock {Perturbative dynamics and relativistic effects of a dyonic
  Kalb-Ramond black hole}.
\newblock 5 2026{\natexlab{a}}.

\bibitem[Fu et~al.(2021)Fu, Zhao, and Liu]{Fu2021}
Qi-Ming Fu, Li~Zhao, and Yu-Xiao Liu.
\newblock {Weak deflection angle by electrically and magnetically charged black
  holes from nonlinear electrodynamics}.
\newblock \emph{Phys. Rev. D}, 104\penalty0 (2):\penalty0 024033, 2021.
\newblock \doi{10.1103/PhysRevD.104.024033}.

\bibitem[Ara{\'u}jo~Filho et~al.(2025{\natexlab{d}})Ara{\'u}jo~Filho, Heidari,
  Lobo, and Bezerra]{AraJCAP2025}
A.~A. Ara{\'u}jo~Filho, N.~Heidari, I.~P. Lobo, and V.~B. Bezerra.
\newblock {Gravitational signatures of a nonlinear electrodynamics in f(R,T)
  gravity}.
\newblock \emph{JCAP}, 09:\penalty0 015, 2025{\natexlab{d}}.
\newblock \doi{10.1088/1475-7516/2025/09/015}.
\newblock [Erratum: JCAP 01, E01 (2026)].

\bibitem[Crisnejo and Gallo(2018)]{CrisnejoGallo2018}
Gabriel Crisnejo and Emanuel Gallo.
\newblock {Weak lensing in a plasma medium and gravitational deflection of
  massive particles using the Gauss-Bonnet theorem. A unified treatment}.
\newblock \emph{Phys. Rev. D}, 97\penalty0 (12):\penalty0 124016, 2018.
\newblock \doi{10.1103/PhysRevD.97.124016}.

\bibitem[Abd El~Dayem et~al.(2024)]{GRAVITY2024}
Karim Abd El~Dayem et~al.
\newblock {Improving constraints on the extended mass distribution in the
  Galactic center with stellar orbits}.
\newblock \emph{Astron. Astrophys.}, 692:\penalty0 A242, 2024.
\newblock \doi{10.1051/0004-6361/202452274}.

\bibitem[Psaltis et~al.(2020)]{Psaltis2020}
Dimitrios Psaltis et~al.
\newblock {Gravitational Test Beyond the First Post-Newtonian Order with the
  Shadow of the M87 Black Hole}.
\newblock \emph{Phys. Rev. Lett.}, 125\penalty0 (14):\penalty0 141104, 2020.
\newblock \doi{10.1103/PhysRevLett.125.141104}.

\bibitem[Bozza and Mancini(2012)]{BozzaMancini2012}
V.~Bozza and L.~Mancini.
\newblock {Observing gravitational lensing effects by Sgr A* with GRAVITY}.
\newblock \emph{Astrophys. J.}, 753:\penalty0 56, 2012.
\newblock \doi{10.1088/0004-637X/753/1/56}.

\bibitem[Gebhardt et~al.(2011)Gebhardt, Adams, Richstone, Lauer, Faber,
  Gultekin, Murphy, and Tremaine]{Gebhardt2011}
Karl Gebhardt, Joshua Adams, Douglas Richstone, Tod~R. Lauer, S.~M. Faber,
  Kayhan Gultekin, Jeremy Murphy, and Scott Tremaine.
\newblock {The Black-Hole Mass in M87 from Gemini/NIFS Adaptive Optics
  Observations}.
\newblock \emph{Astrophys. J.}, 729:\penalty0 119, 2011.
\newblock \doi{10.1088/0004-637X/729/2/119}.

\bibitem[{GRAVITY Collaboration}(2017)]{GRAVITY2017}
{GRAVITY Collaboration}.
\newblock {First Light for GRAVITY: Phase Referencing Optical Interferometry
  for the Very Large Telescope Interferometer}.
\newblock \emph{Astronomy \& Astrophysics}, 602:\penalty0 A94, 2017.
\newblock \doi{10.1051/0004-6361/201730838}.

\bibitem[Vagnozzi et~al.(2023)]{Vagnozzi2023}
Sunny Vagnozzi et~al.
\newblock {Horizon-scale tests of gravity theories and fundamental physics from
  the Event Horizon Telescope image of Sagittarius A$^*$}.
\newblock \emph{Class. Quant. Grav.}, 40\penalty0 (16):\penalty0 165007, 2023.
\newblock \doi{10.1088/1361-6382/acd97b}.

\bibitem[Akiyama et~al.(2024)]{EHTM872024}
Kazunori Akiyama et~al.
\newblock {The persistent shadow of the supermassive black hole of M~87. I.
  Observations, calibration, imaging, and analysis}.
\newblock \emph{Astron. Astrophys.}, 681:\penalty0 A79, 2024.
\newblock \doi{10.1051/0004-6361/202347932}.

\bibitem[Akiyama et~al.(2025)]{EHTM872025}
Kazunori Akiyama et~al.
\newblock {Horizon-scale variability of M87* from 2017{\textendash}2021 EHT
  observations}.
\newblock \emph{Astron. Astrophys.}, 704:\penalty0 A91, 2025.
\newblock \doi{10.1051/0004-6361/202555855}.

\bibitem[Ara{\'u}jo~Filho et~al.(2026{\natexlab{b}})Ara{\'u}jo~Filho, Heidari,
  Lobo, and Bezerra]{AraBumblebee2026}
A.~A. Ara{\'u}jo~Filho, N.~Heidari, Iarley~P. Lobo, and V.~B. Bezerra.
\newblock {Gravitational aspects of a new bumblebee black hole}.
\newblock \emph{Annals Phys.}, 489:\penalty0 170469, 2026{\natexlab{b}}.
\newblock \doi{10.1016/j.aop.2026.170469}.

\bibitem[Tsukamoto(2023)]{Tsukamoto2023}
Naoki Tsukamoto.
\newblock {Gravitational lensing by using the 0th order of affine perturbation
  series of the deflection angle of a ray near a photon sphere}.
\newblock \emph{Eur. Phys. J. C}, 83\penalty0 (4):\penalty0 284, 2023.
\newblock \doi{10.1140/epjc/s10052-023-11419-9}.

\bibitem[Stefanov et~al.(2010)Stefanov, Yazadjiev, and Gyulchev]{Stefanov2010}
Ivan~Zh. Stefanov, Stoytcho~S. Yazadjiev, and Galin~G. Gyulchev.
\newblock {Connection between Black-Hole Quasinormal Modes and Lensing in the
  Strong Deflection Limit}.
\newblock \emph{Phys. Rev. Lett.}, 104:\penalty0 251103, 2010.
\newblock \doi{10.1103/PhysRevLett.104.251103}.

\bibitem[Salehi et~al.(2025)Salehi, Kumar~Walia, Chang, and
  Kocherlakota]{Salehi2025}
Kiana Salehi, Rahul Kumar~Walia, Dominic~O. Chang, and Prashant Kocherlakota.
\newblock {Influence of observer{\textquoteright}s inclination and spacetime
  structure on photon ring observables}.
\newblock \emph{Phys. Rev. D}, 111\penalty0 (10):\penalty0 104057, 2025.
\newblock \doi{10.1103/PhysRevD.111.104057}.

\bibitem[Lockhart and Gralla(2022)]{LockhartGralla2022}
Will Lockhart and Samuel~E. Gralla.
\newblock {How narrow is the M87* ring {\textendash} II. A new geometric
  model}.
\newblock \emph{Mon. Not. Roy. Astron. Soc.}, 517\penalty0 (2):\penalty0
  2462--2470, 2022.
\newblock \doi{10.1093/mnras/stac2743}.

\bibitem[Ara{\'u}jo~Filho(2025{\natexlab{a}})]{AraujoFilho:2025rwr}
A.~A. Ara{\'u}jo~Filho.
\newblock {Particle production induced by a Lorentzian non-commutative
  spacetime}.
\newblock \emph{Annals Phys.}, 481:\penalty0 170167, 2025{\natexlab{a}}.
\newblock \doi{10.1016/j.aop.2025.170167}.

\bibitem[Ara{\'u}jo~Filho(2025{\natexlab{b}})]{AraujoFilho:2025hkm}
A.~A. Ara{\'u}jo~Filho.
\newblock {How does non-metricity affect particle creation and evaporation in
  bumblebee gravity?}
\newblock \emph{JCAP}, 06:\penalty0 026, 2025{\natexlab{b}}.
\newblock \doi{10.1088/1475-7516/2025/06/026}.
\newblock [Erratum: JCAP 02, E01 (2026)].

\bibitem[Ara{\'u}jo~Filho(2025{\natexlab{c}})]{AraujoFilho:2024ctw}
A.~A. Ara{\'u}jo~Filho.
\newblock {Particle creation and evaporation in Kalb-Ramond gravity}.
\newblock \emph{JCAP}, 04:\penalty0 076, 2025{\natexlab{c}}.
\newblock \doi{10.1088/1475-7516/2025/04/076}.

\bibitem[de~Paula et~al.(2025)de~Paula, Leite, and Crispino]{dePaula:2025kif}
Marco A.~A. de~Paula, Luiz C.~S. Leite, and Lu{\'\i}s C.~B. Crispino.
\newblock {Sufficient conditions for unbounded superradiance in black hole
  spacetimes sourced by nonlinear electrodynamics}.
\newblock \emph{Phys. Rev. D}, 111\penalty0 (10):\penalty0 104010, 2025.
\newblock \doi{10.1103/PhysRevD.111.104010}.

\bibitem[Dolan et~al.(2024)Dolan, de~Paula, Leite, and Crispino]{Dolan:2024qqr}
Sam~R. Dolan, Marco A.~A. de~Paula, Luiz C.~S. Leite, and Lu{\'\i}s C.~B.
  Crispino.
\newblock {Superradiant instability of a charged regular black hole}.
\newblock \emph{Phys. Rev. D}, 109\penalty0 (12):\penalty0 124037, 2024.
\newblock \doi{10.1103/PhysRevD.109.124037}.

\bibitem[de~Paula et~al.(2024)de~Paula, Leite, Dolan, and
  Crispino]{dePaula:2024xnd}
Marco A.~A. de~Paula, Luiz C.~S. Leite, Sam~R. Dolan, and Lu{\'\i}s C.~B.
  Crispino.
\newblock {Absorption and unbounded superradiance in a static regular black
  hole spacetime}.
\newblock \emph{Phys. Rev. D}, 109\penalty0 (6):\penalty0 064053, 2024.
\newblock \doi{10.1103/PhysRevD.109.064053}.

\bibitem[Furtado et~al.(2023)Furtado, Hassanabadi, Reis,
  et~al.]{furtado2023thermal}
J~Furtado, H~Hassanabadi, JAAS Reis, et~al.
\newblock Thermal analysis of photon-like particles in rainbow gravity.
\newblock \emph{arXiv preprint arXiv:2305.08587}, 2023.

\bibitem[Ara{\'u}jo~Filho and Petrov(2021)]{araujo2021bouncing}
A.~A Ara{\'u}jo~Filho and A.~Yu Petrov.
\newblock Bouncing universe in a heat bath.
\newblock \emph{International Journal of Modern Physics A}, 36\penalty0
  (34n35):\penalty0 2150242, 2021.

\bibitem[Ara{\'u}jo~Filho(2025{\natexlab{d}})]{AraujoFilho:2025fwd}
A.~A. Ara{\'u}jo~Filho.
\newblock {Particle motion and thermal effects around a Kalb{\textendash}Ramond
  black hole}.
\newblock \emph{Eur. Phys. J. C}, 85\penalty0 (9):\penalty0 1002,
  2025{\natexlab{d}}.
\newblock \doi{10.1140/epjc/s10052-025-14752-3}.

\bibitem[Ara{\'u}jo~Filho(2026{\natexlab{b}})]{AraujoFilho:2026yaj}
A.~A. Ara{\'u}jo~Filho.
\newblock {Non-metricity effects on electron scattering in bumblebee gravity}.
\newblock \emph{Eur. Phys. J. C}, 86\penalty0 (7):\penalty0 767,
  2026{\natexlab{b}}.
\newblock \doi{10.1140/epjc/s10052-026-16038-8}.

\bibitem[Ara{\'u}jo~Filho(2022)]{araujo2022particles}
A.~A Ara{\'u}jo~Filho.
\newblock Particles in loop quantum gravity formalism: a thermodynamical
  description.
\newblock \emph{Annalen der Physik}, 534\penalty0 (12):\penalty0 2200383, 2022.

\end{thebibliography}

\end{document}